\documentclass{article}

\usepackage{arxiv}

\usepackage[utf8]{inputenc} 
\usepackage[T1]{fontenc}    
\usepackage{hyperref}       
\usepackage{url}            
\usepackage{booktabs}       
\usepackage{amsfonts}       
\usepackage{nicefrac}       
\usepackage{microtype}      
\usepackage{lipsum}		
\usepackage{graphicx}
\usepackage{doi}

\usepackage{subcaption}

\usepackage{color}

\usepackage{graphics} 
\usepackage{epsfig} 
\usepackage{amsmath} 
\usepackage{amssymb}  
\usepackage{breqn}

\usepackage[backend=biber,sorting=none]{biblatex} %
\newtheorem{theorem}{Theorem}
\newtheorem{definition}{Definition}

\title{Sparse Polynomial Optimisation for Neural Network Verification}

\author{Matthew Newton \\
	Department of Engineering Science\\
	University of Oxford\\
	Oxford, OX1 3PJ, UK \\
	\texttt{matthew.newton@eng.ox.ac.uk} \\
	
	\And
	Antonis Papachristodoulou\\
	Department of Engineering Science\\
	University of Oxford\\
	Oxford, OX1 3PJ, UK \\
	\texttt{antonis@eng.ox.ac.uk} \\
}

\date{}

\renewcommand{\headeright}{}
\renewcommand{\undertitle}{}
\renewcommand{\shorttitle}{Neural Networks and Sparse Polynomial Optimisation}

\begin{document}
\maketitle

\begin{abstract} 
The prevalence of neural networks in society is expanding at an increasing rate. It is becoming clear that providing robust guarantees on systems that use neural networks is very important, especially in safety-critical applications. A trained neural network's sensitivity to adversarial attacks is one of its greatest shortcomings. To provide robust guarantees, one popular method that has seen success is to bound the activation functions using equality and inequality constraints. However, there are numerous ways to form these bounds, providing a trade-off between conservativeness and complexity. Depending on the complexity of these bounds, the computational time of the optimisation problem varies, with longer solve times often leading to tighter bounds. We approach the problem from a different perspective, using sparse polynomial optimisation theory and the Positivstellensatz, which derives from the field of real algebraic geometry. The former exploits the natural cascading structure of the neural network using ideas from chordal sparsity while the later asserts the emptiness of a semi-algebraic set with a nested family of tests of non-decreasing accuracy to provide tight bounds. We show that bounds can be tightened significantly, whilst the computational time remains reasonable. We compare the solve times of different solvers and show how the accuracy can be improved at the expense of increased computation time. We show that by using this sparse polynomial framework the solve time and accuracy can be improved over other methods for neural network verification with ReLU, sigmoid and tanh activation functions. 
\end{abstract}


\section{Introduction}
The field of machine learning has seen a huge resurgence of interest over the past decade. This is primarily down to the large increase of research into neural networks. In particular, the development of Alexnet \cite{akriz} and Resnet \cite{resnet} showed a step increase in the capacity of neural networks to perform complex tasks that were once thought to be impossible to compute by machine. The increase in computational power available and the abundance of big-data, has lead to an increase in industrial applications of neural networks and their prevalence is ever expanding. Key examples of these areas include but are not limited to image recognition, weather prediction and natural language processing \cite{czhang}, \cite{tbro}. 

With the success of neural networks in general application areas, the transition into safety-critical applications is an important consideration, such as autonomous vehicle technology. Neural networks provide the opportunity to bridge the gap to make a once-thought impossible task an actuality. However, before this can be achieved, the research community along with industry must overcome one of the biggest shortcomings of this new technology, which is the neural network's sensitivity to adversarial inputs, where large changes in the output set can be caused by relatively small changes in the input set. There has been a considerable amount of effort to better our understanding of neural networks and to provide certificates on such systems. However, there are still significant hurdles to pass for their wide-spread use in safety-critical applications. The traditional `black box' approach for systems of this type is not sufficient in this case. 

In the field of control theory, there exists work in the area of neural networks dating back to the 1990s \cite{wmiller}. There has been a sparked interest in this area along with the emergence of the parallel field of reinforcement learning, which mostly focuses on a purely data-based approach to control by considering an agent in an environment. Deep reinforcement learning has harnessed the power of neural networks and enabled a decision making agent to greatly outperform humans in many complex tasks, such as the video game Dota 2 and the board game Go \cite{alphago}. Bounds to quantify their safety have been developed, however are often overly conservative and do not quantify the performance of the algorithm sufficiently \cite{rsut}. The success of the reinforcement learning community combined with new advancements in robust control methods, motivates work at their intersection. Diverging away from traditional model-based approaches within the scope of control theory has the potential to lead to many exciting developments. Examples of recent work that has shown how neural networks can be used in control systems include \cite{sdut}, that uses a neural network to learn and verify a feedback control system; and \cite{spot} that can identify false data injection attacks in control systems by implementing an artificial neural network. Another aspect that has been focused on is using a neural network to show stability of feedback systems by learning Lyapunov functions \cite{schen}, \cite{aabat}.

There are many methods to compute robust guarantees on a neural network to verify its safety. One approach focuses in finding the Lipchitz constant for the neural network~\cite{mfaz2}. Moreover, in \cite{twen}, the authors use Extreme Value Theory to better quantify the robustness. These Lipchitz constants can then be used to train the neural network to be robust \cite{ppau}. Another method is to use Satisfiability Modulo Theory to quantify the safety of a neural network \cite{xhua}. There have been competitions such as ARCH-COMP20 \cite{dlop} that were organised to compare different methods, where participants were asked to certify control systems. Another common approach of achieving this is by placing bounds on the activation functions that are often non-linear on each node in the neural network \cite{hsal}: this is the idea that is used in this paper. Due to the large number of formulations and choices of activation functions, there is a sizeable literature showing this method to be successful, with each result aiming to find tighter and more efficient bounds on the properties of the neural network. The simplest methods are zero-th order methods such as interval bound propagation \cite{sgow}, that computes the worst-case scenarios out of each layer in the network. Another comparable method is CROWN, that uses similar ideas with a sophisticated implementation \cite{hzha}, which often provides better results. Frameworks that provide linear bounds on the activation functions will result in linear programs, which can be solved using various optimisation methods \cite{ddeep}, \cite{rbun}. There are many ways that these linear bounds can be imposed for different activation functions. On top of this, researchers have found other methods to improve the accuracy and scalability of the problem. By using a dual approach, \cite{kdvi} focuses on improving the scalability issue. \cite{gsin} proposes a new parametric framework called `k-ReLU' that combines the constraints from multiple activation functions. By considering the multi-variate input space on the activation functions, \cite{ctja} results in a similar method. A scalable approach that uses this linear programming framework is DeepSplit that uses an operator splitting method to find the bounds \cite{schen2}.

To improve the tightness of the linear relaxations, \cite{arag} introduces semidefinite relaxations for the certification of the robustness properties of neural networks. Another semidefinite programming approach that provides tight bounds on the neural network is built on the formulation of quadratic constraints \cite{mfaz}. This type of formulation can also be used on feedback control systems to conduct reachability analysis \cite{hhu}, another approach shows that these quadratic constraints can be extended to integral quadratic constraints to analyse the stability of a neural network controller \cite{hyin}. However, the semidefinite programming approach has the drawback of scaling worse than linear programming methods. \cite{mnew} improves the scalability of this framework by exploiting the sparsity pattern which is shown to match the intrinsic structure of the neural network. Another method to overcome this is to use an iterative eigenvector approach to improve the efficiency of large scale instances of neural networks \cite{sdat}. There are additional methods that use ideas from robust control theory, for example \cite{ywan} utilises these tools to certify a neural network control policy. Another approach is using output range analysis to find these certificates \cite{bkar}. By treating the neural network as a dynamical system, \cite{sdek} are able to obtain performance metrics on adversarial inputs. \cite{ttan} uses notions of stability in dynamical systems theory to diagnose and prevent instability in recurrent video processing. As mentioned above, Lipchitz constants are a popular metric to quantify the sensitivity of a neural network, and \cite{mfaz2} proposes a method of obtaining them using semidefinite programming. This approach is extended using incremental quadratic constraints in \cite{nhas}. Lipchitz bounds have been shown to be computed in a scalable way, as \cite{flat} uses sparse polynomial optimisation to achieve this. The idea of Lipchitz constants have been applied to equilibrium neural networks, which are a general class of neural networks \cite{mrev}. 

\subsection{Our Contribution}
There is a large amount of prior work that has focused on obtaining tighter bounds on the potential outputs of a neural network or improving the scalability of the optimisation problem. However, previous works have not tried to tackle these problems simultaneously. Another important drawback of the previous literature is that each framework is analysed in isolation, with most individual papers focusing on one type of constraint for an individual activation function. It is of course important to optimise bounds for each activation function to provide the best results. However, having a unified framework to certify the bounds on the neural network would be useful, so that these methods can be combined together to give the best possible bounds in all scenarios. We therefore look for approaches that will both generalise the bounds on the neural network, whilst considering the scalability of solving the optimisation problem. 
\begin{itemize}
    \item We propose the problem in a sparse polynomial optimisation framework. To do this we use a theorem from real algebraic geometry called the Positivstellensatz, which certifies the emptiness of a semi-algebraic set using algebra. 
    \item We formulate the neural network verification problem as a set of equality and inequality constraints, which results in a semi-algebraic set. We then search for the emptiness of this set by using the algebraic formulation to find the bounds on the neural network's output, using Sum of Squares and semidefinite programming methods. The Postivstellensatz can be altered to make it more or less conservative meaning that we can easily trade off solution accuracy and computational complexity. 
    \item We use the observation that a neural network possesses a natural cascading structure to reduce the semidefinite programming constraints to smaller constraints. The key idea that underpins this method is a decomposition theorem that explains the link between chordal graphs and positive semidefinite matrices, showing that a large positive semidefinite constraint can be split up into smaller positive semidefinite constraints. We analyse the optimisation framework using theory from sparse polynomial optimisation and show how the variables are separated into different constraints.
    \item We implement the optimisation problem and show through numerous examples that this method can be used to improve both the computational time and accuracy. The examples we show have a varying number of nodes and layers for ReLU, sigmoid and tanh activation functions. Previously, neural networks of this size have not been verified to this degree of accuracy.
\end{itemize}

In Section \ref{sec:NN} we provide an overview of the neural network verification problem and how the input-output relationships can be represented as a series of inequality and equality constraints. The Positivstellensatz is introduced in Section \ref{sec:problem} and we then describe how it can be used in the neural network verification problem. The theory behind sparse polynomial optimisation and the link to Positivstellensatz is delineated in Section \ref{sec:sparsePOP}. The theory is then applied to the specific neural network verification problem in Section \ref{sec:sparseNN}. In section \ref{sec:results} the results from various experiments are presented and analysed. The paper is then concluded in Section \ref{sec:conc}, outlining plans for future work. 

\section{Neural Network Verification} \label{sec:NN}
\begin{figure}[h] 
    \centering 
    \includegraphics[height=5cm]{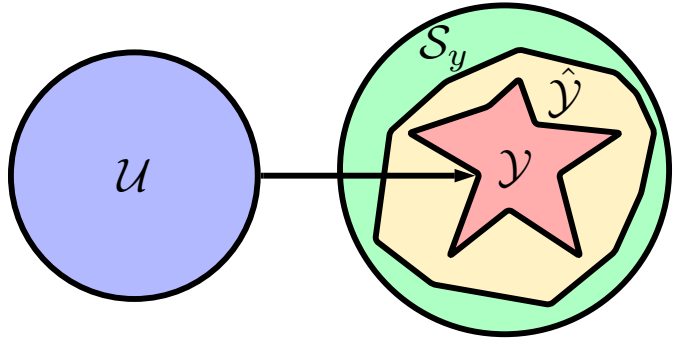}
    \caption{Diagram showing the neural network verification problem. $\mathcal{U}$ represents the input set, $\mathcal{Y}$ represents the output set and $\hat{\mathcal{Y}}$ is the approximation of this set. The safe set of outputs is denoted as $\mathcal{S}_{y}$.}  \label{fig:NNVerification}
\end{figure}

The set of real $n$-dimensional vectors is denoted by $\mathbb{R}^{n}$ and the set of $m \times n$-dimensional matrices is denoted by $\mathbb{R}^{m \times n}$. A multi-layer feed-forward neural network can be described as a non-linear function $\pi: \mathbb{R}^{n_{u}} \rightarrow \mathbb{R}^{n_{y}}$, where $n_{u}$ is the number of inputs and $n_{y}$ is the number of outputs. Consider the set of all possible inputs into the neural network $\mathcal{U} \subset \mathbb{R}^{n_{u}}$; the neural network will map these inputs to a set of outputs $\mathcal{Y} \subset \mathbb{R}^{n_{y}}$. This mapping can be expressed as
\begin{equation*}
    \mathcal{Y} = \pi(\mathcal{U}) := \{ y \in \mathbb{R}^{n_{y}} \: | \: y = \pi(u),\: u \in \mathcal{U} \}.
\end{equation*}
To certify the neural network we ask that the outputs of the neural network lie within a safe region $\mathcal{S}_{y}$ given a set of inputs $\mathcal{U}$. This is the principle of the neural network verification problem. Conversely, the safe set of inputs are defined as $\mathcal{S}_{u} := \pi^{-1}( \mathcal{S}_{y})$. Note that there are no guarantees on the convexity of the $\mathcal{S}_{y}$ set; in fact it is likely that it is non-convex. It is therefore computationally expensive to check if these outputs lie in a safe set. To overcome this issue a relaxation can be computed as a conservative approximation of the set $\mathcal{Y}$, denoted $\hat{\mathcal{Y}}$. We now instead check the condition  $\hat{\mathcal{Y}} \subseteq \mathcal{S}_{y}$ to verify the neural network. A diagram of this setup is shown in Figure \ref{fig:NNVerification}.

\subsection{Neural Network Model}
We consider a feed-forward fully connected neural network $\pi: \mathbb{R}^{n_{u}} \rightarrow \mathbb{R}^{n_{y}}$, with $\ell$ layers. This can be represented by the set of equations:
\begin{eqnarray*} \label{eq:nn1}
    x^{0} &=& u, \\ 
    v^{k} &=& W^{k}x^{k} + b^{k}, \: \mathrm{for} \: k = 0,\dots, \ell - 1, \\ 
    x^{k+1} &=& \phi (v^{k}), \: \mathrm{for} \: k = 0,\dots, \ell - 1, \\ 
    \pi(u) &=& W^{\ell}x^{\ell} + b^{\ell}, 
\end{eqnarray*} 
where $W^{k} \in \mathbb{R}^{n_{k+1} \times n_{k}}$, $b^{k} \in \mathbb{R}^{n_{k+1}}$ are the weights matrix and biases of the $(k+1)^{th}$ layer respectively and $u = x^{0} \in \mathbb{R}^{n_{u}}$ is the input into the network. The number of neurons in the $k^{th}$ layer is denoted by $n_{k}$; the total number of neurons in the neural network is therefore $n = \sum_{k=1}^{\ell} n_{k}$. The non-linear activation function $\phi$ is applied element-wise to the $v^{k} = W^{k}x^{k} + b^{k}$ terms such that
\begin{equation*}
    \phi(v^{k}) := [\phi(v_{1}^{k}),\dots, \phi(v_{n_{k}}^{k})]^{T}, \: v^{k} \in \mathbb{R}^{n_{k}}, 
\end{equation*}
where $\phi$ is the activation function and $v_{j}^{k}$ is the pre-activation value. There are many different types of activation functions such as ReLU, tanh, sigmoid, GELU, softsign and ELU \cite{cnwa}.

\subsection{Problem Overview} 
We now describe how constraints on the non-linear activation functions can be created. How these constraints are used subsequently to form an optimisation problem to verify a neural network is presented in Section \ref{sec:problem}. The constraints can be split into three main categories, by considering the input, hidden layer and output constraints separately. 

We assume that the input constraints are bounded by a hyper-rectangle defined by $\mathcal{U} = \{ u \in \mathbb{R}^{n_{u}} \: | \: \underbar{$u$} \leq u \leq \overline{u} \}$. Therefore, the input constraints can be written as
\begin{equation*}
    g_{in}^{1} = u - \underbar{$u$} \geq 0, \: g_{in}^{2} = -u + \overline{u} \geq 0.
\end{equation*}
We assume that the safe set can be represented by the polytope
\begin{equation*}
    \mathcal{S}_{y} = \bigcap_{m=1}^{M} \{ y \in \mathbb{R}^{n_{y}}\: | \: c_{m}^{T}y - d_{m} \leq 0 \},
\end{equation*}
where $c_m\in \mathbb{R}^{n_{y}}$ and $d_m \in \mathbb{R}$ are given parameters of the polytope. 

In the optimisation framework the faces of the polytope are considered separately, where we attempt to obtain bounds on the elements $d_m$ to reduce the volume of the polytope. Therefore, the search for bounds on the safe output set can be split into $M$ optimisation programs. The output constraints contain a decision variable that can be minimized in the optimisation program such that 
\begin{equation*}
    g_{out}^{m} = \gamma_{m} - c_{m}^{T}y \geq 0,
\end{equation*}
where $\gamma_{m}$ is the decision variable to be optimised and $m$ denotes the $m^{th}$ optimisation program.

The hidden layer constraints can be represented by relationships between $\phi(v^{k})$ and $v^{k}$, which can be expressed through properties of the activation function. Since the activation functions are applied element-wise to the $v^{k}$ terms, we consider the relationship between $\phi(v_{j}^{k})$ and $v_{j}^{k}$ separately. To simplify the notation, we denote the output of the activation function $\phi(v_{j}^{k})$ as $\phi$, and the input to the activation function $v_{j}^{k}$ as $x$. These relationships can be formed through many means; popular methods include sector, slope and box constraints, however there is no limitation of how these constraints can be expressed. It is also possible to compute a conservative approximation of the bounds using an efficient pre-processing step known as interval bound propagation (IBP) \cite{sgow}. IBP is a zero-th order method, which uses interval arithmetic to find the minimum and maximum bounds on the activation function, $(\underbar{$\phi$}, \overline{\phi})$ such that 
\begin{equation} \label{eq:ibp}
    \phi - \underbar{$\phi$} \geq 0, \: -\phi + \overline{\phi} \geq 0 \textrm{  everywhere.}
\end{equation}
The pre-activation values from IBP are denoted as $(\underbar{$x$},\overline{x})$. There are other efficient methods that can be used in this pre-processing step such as CROWN \cite{hzha} that work for general activation functions and can provide better results, but take longer to compute. These box constraints are used in the optimisation framework and there are many different possible constraints that can be used and combined with the preprocessing values. There are a few classes of constraints that can be categorised, we will outline these now.

\subsection{Sector Constraints}
A common trait of an activation function is that it is monotonically increasing - a function that has this property can often be bounded by a sector constraint. 
\begin{definition}
Consider the non-linear activation function $\phi(x)$ with $\phi(0) = \lambda$. A sector constraint states that $\phi(x)$ lies in the sector $[\alpha, \beta] \: (\alpha \leq \beta < \infty)$ if the following condition holds
\begin{equation*}
    (\phi(x) - (\alpha x + \lambda))((\beta x + \lambda) - \phi(x)) \geq 0, \: \forall \: x \in \mathbb{R}.
\end{equation*}
\end{definition}
Note that a sector constraint only considers the relationship between the activation function's output and input. We show in Section \ref{sec:sparseNN} that this point is important to preserve the sparsity property in the algorithm formulation.

\subsection{Slope Constraints}
By considering the slope of two activation functions from different layers, another set of bounds that can be used are known as slope constraints \cite{mfaz}. Since the activation functions have a predefined structure, their slopes are often bounded between two values, which can be used to form a sector constraint such that
\begin{equation*}
    \alpha \leq \frac{\phi(x_{2}) - \phi(x_{1})}{x_{2} - x_{1}} \leq \beta.
\end{equation*}
Therefore, any two nodes in the neural network must satisfy
\begin{equation*}
    (\phi(x_{i}) - \phi(x_{j}) - \alpha (x_{i} - x_{j}))( \beta (x_{i} - x_{j}) - (\phi(x_{i}) - \phi(x_{j})) \geq 0,  
\end{equation*}
$\forall i,j = 1, \dots, n, \: i \neq j$. For example for the ReLU and tanh activation functions the slope restricted sectors are $\alpha = 0, \beta = 1$ and for sigmoid the values are $\alpha = 0, \beta = 0.25$. However, the big issue with constraints of this type is the lack of scalability. As the number of neurons in the network increases, the number of constraints increases as $n \choose 2$. In addition to this, the constraints are not always a function of the activation functions in two consecutive layers, which will destroy the sparsity in the algorithm formulation. Therefore, they will not be considered further in this paper.

\subsection{ReLU Function}
\begin{figure*}
    \centering
    \begin{subfigure}[b]{0.49\textwidth}
        \centering
        \includegraphics[width=\textwidth]{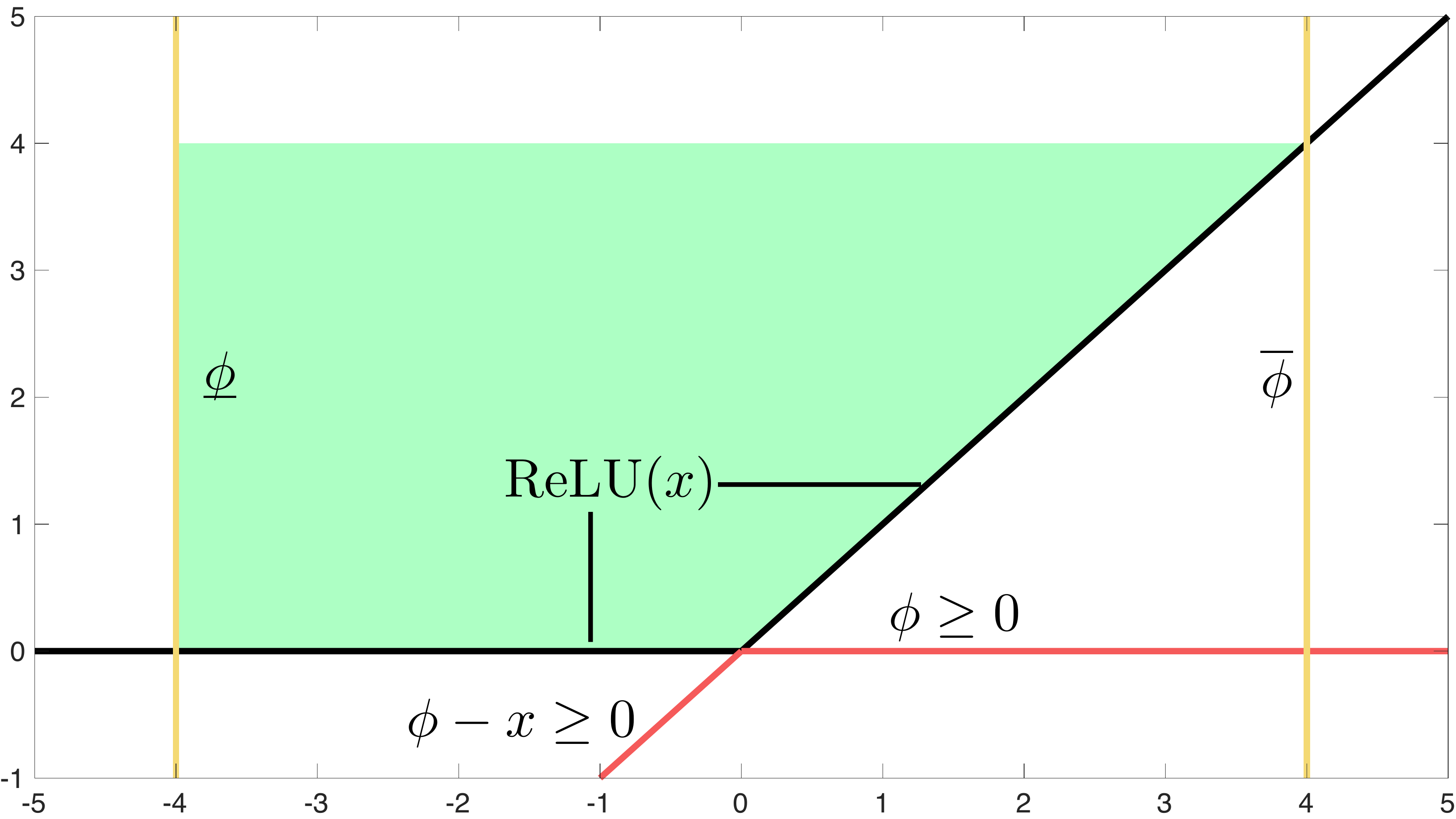}
        \caption{The inequality constraints in \eqref{eq:relu}.}
        \label{fig:relu1}
    \end{subfigure}
    \hfill
    \begin{subfigure}[b]{0.49\textwidth}  
        \centering 
        \includegraphics[width=\textwidth]{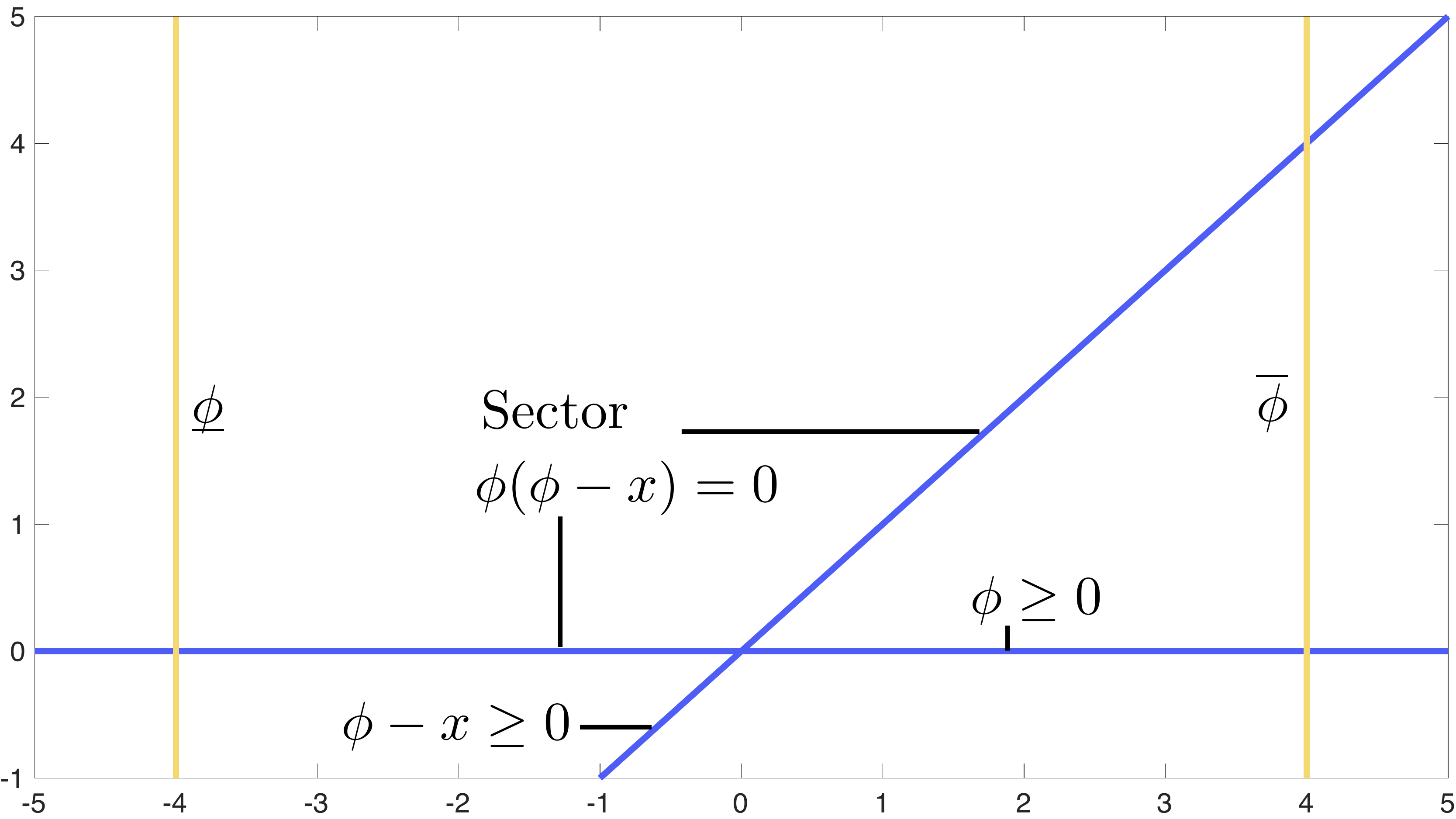}
        \caption{The equality constraints in \eqref{eq:relu}.} 
        \label{fig:relu2}
    \end{subfigure}
    \vskip\baselineskip
    \begin{subfigure}[b]{0.49\textwidth}   
        \centering 
        \includegraphics[width=\textwidth]{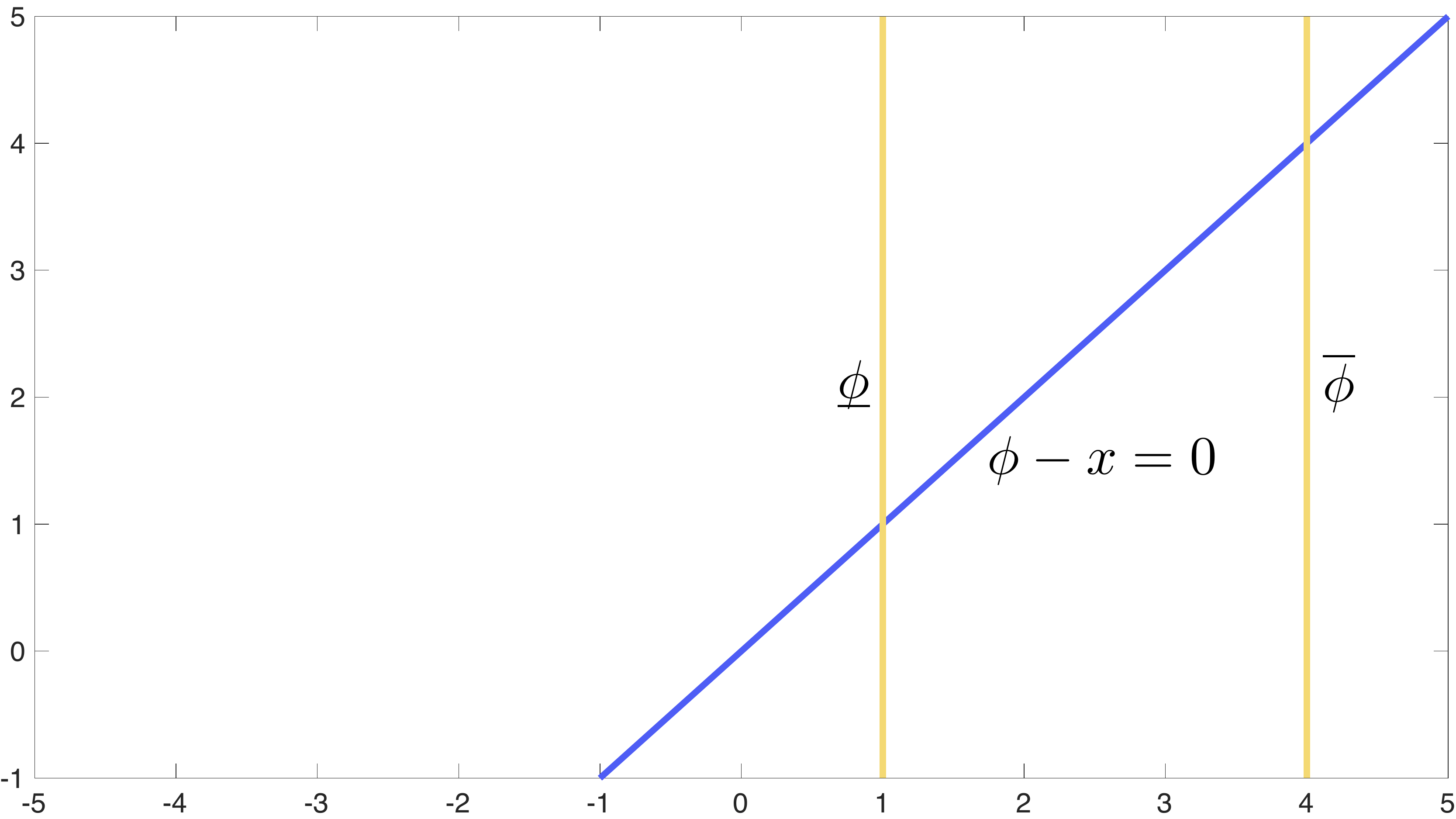}
        \caption{The equality constraint in \eqref{eq:relu} when $\underbar{$\phi$} > 0$.}
        \label{fig:relu3}
    \end{subfigure}
    \hfill
    \begin{subfigure}[b]{0.49\textwidth}   
        \centering 
        \includegraphics[width=\textwidth]{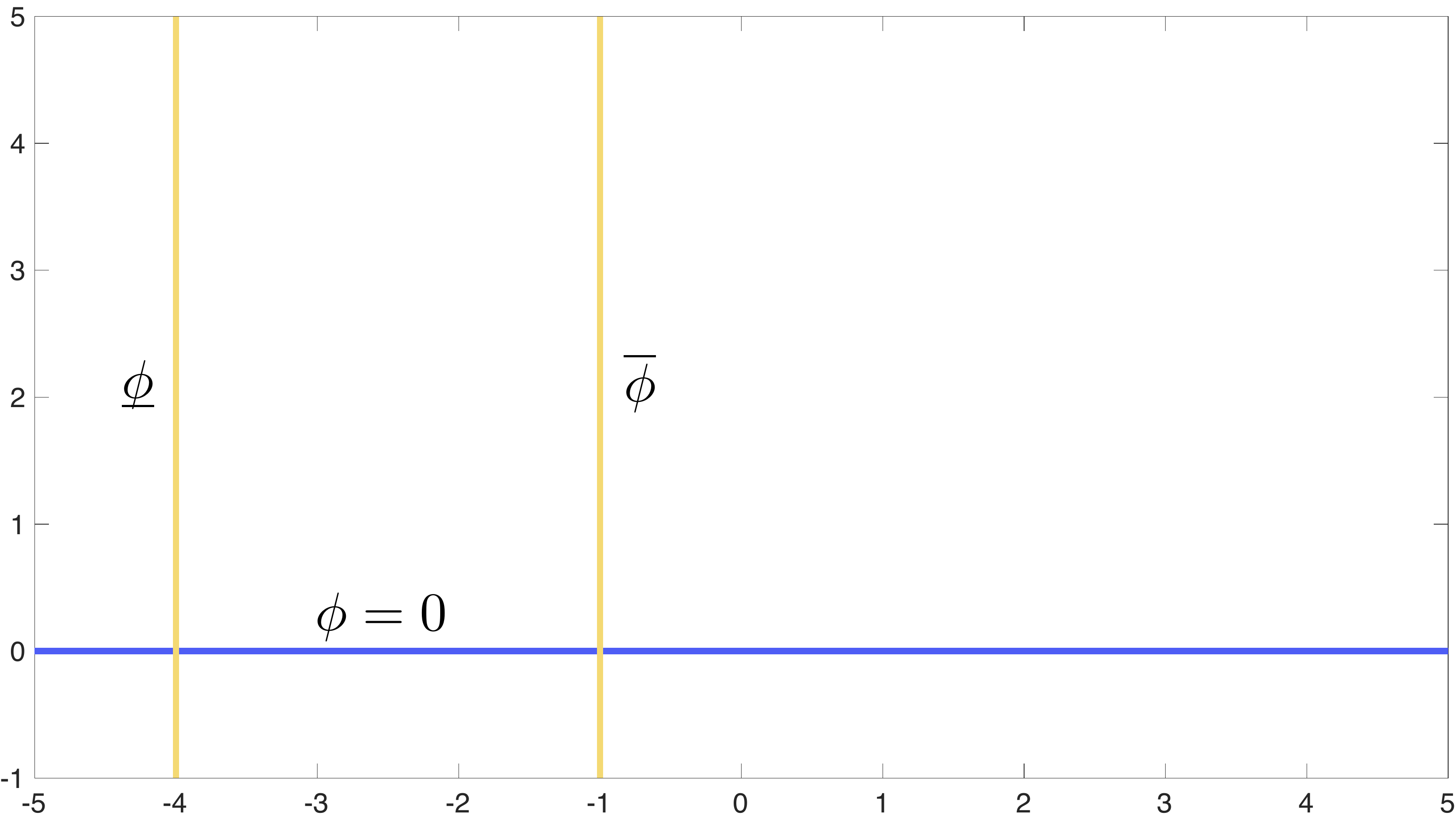}
        \caption{The equality constraint in \eqref{eq:relu} when $\overline{\phi} \leq 0$.}  
        \label{fig:relu4}
    \end{subfigure}
    \caption{Plot showing the inequality constraints in \eqref{eq:relu} that bound the ReLU function (black). The red lines represents the inequality constraints, the blue lines represents the quadratic equality constraint and the yellow lines represent the constraints from the IBP values. The green shaded area is the region bounded by the constraints.}
    \label{fig:relutotal}
\end{figure*}

We will now outline how bounds can be computed for some example activation functions. ReLU is a commonly used function and is given by
\begin{equation*}
    \mathrm{ReLU}(x) = \phi(x) = \begin{cases} 
      0 & x\leq 0, \\
      x & x > 0 .
   \end{cases}
\end{equation*}
We can gain tight bounds on the ReLU function using two inequalities and one equality constraint \cite{arag} such that
\begin{equation} \label{eq:relu}
    \phi \geq 0, \: \phi - x \geq 0, \: \phi(\phi - x) = 0.
\end{equation}
The values from the IBP can be used to further tighten these constraints: if $\overline{\phi} \leq 0$, then we can replace the first inequality constraint with an equality constraint such that $\phi = 0$. Equivalently if $\underbar{$\phi$} > 0$, we can replace the second inequality constraint with $\phi - x = 0$. These constraints are shown visually in Figure \ref{fig:relu1} -- \ref{fig:relu4}. 

\subsection{Sigmoid Activation Function} \label{sec:sig}
The sigmoid function is given by
\begin{equation*}
    \mathrm{sig}(x) = \phi(x) = \frac{1}{1 + e^{-x}}
\end{equation*}
and it can be bounded by a single sector constraint such that 
\begin{equation} \label{eq:singlesec}
    (\phi - 0.5)(0.25x + 0.5 - \phi) \geq 0. 
\end{equation}
However, this bound is very conservative as there is a large uncertainty in the value of the activation function. As shown in \cite{mnew2} this can be tightened drastically by using two sector constraints that are carefully positioned as in Figure \ref{fig:sigsec}. Having two sectors instead of one allows us to capture the point of inflection of the sigmoid function. However, the drawback of this is that there will be twice the number of inequality constraints which will make the optimisation problem more expensive. For details of how the sectors are created, the reader may refer to \cite{mnew2}.

\subsection{Tanh Activation Function}
\begin{figure}[h]
\centering
\begin{subfigure}{\linewidth}
\centering 
   \includegraphics[height=5.75cm]{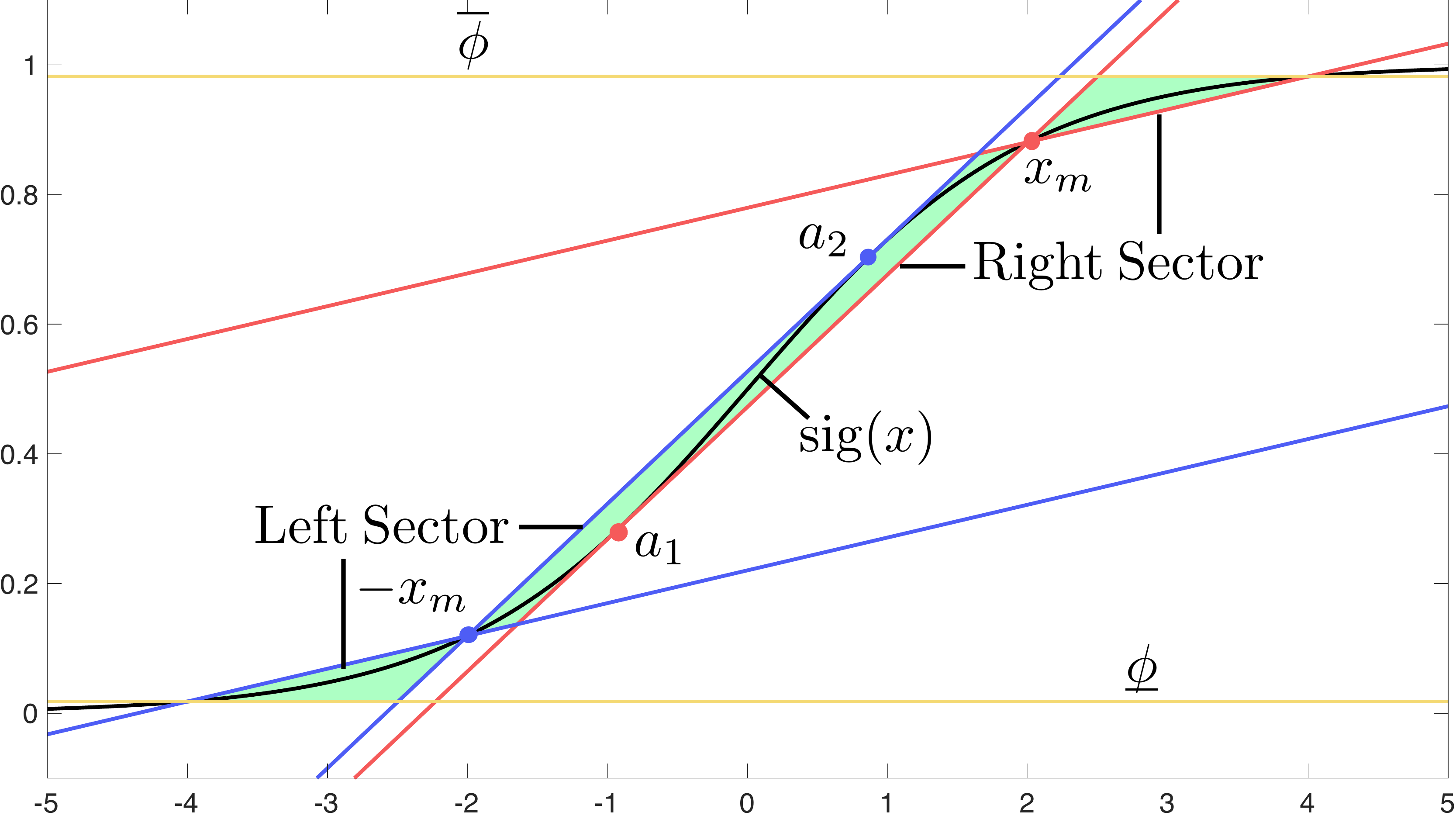}
   \caption{Sigmoid Activation Function}
   \label{fig:sigsec} 
\end{subfigure}\\[1ex]
\begin{subfigure}{\linewidth}
\centering
   \includegraphics[height=5.75cm]{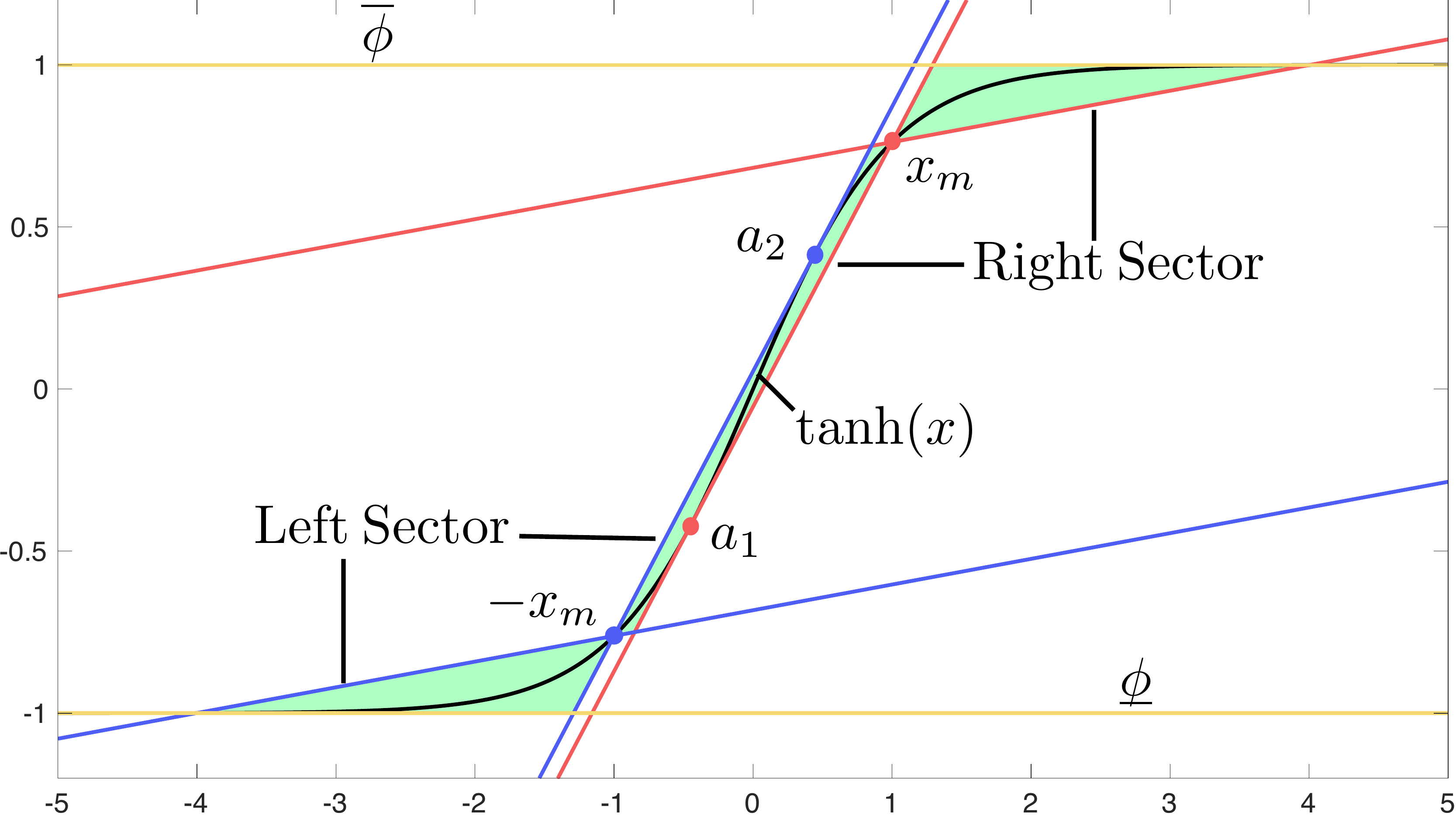}
   \caption{Tanh Activation Function}
   \label{fig:tanhsec}
\end{subfigure}
\caption{Plot showing the two sector constraints (red and blue) that bound the activation function (black). The area in green represents the region defined by the constraints. The point $x_{m}$ is a hyper-parameter to be chosen and defines where the two lines intersect to form the right sector. The lower line passes through the intersection of $\overline{\phi}$ and $\phi(x)$ and the upper line intersects the point $a_{1}$ such that it is tangential to the $\phi(x)$ curve. The same process is repeated with the left sector with mid point of $-x_{m}$.}
\end{figure}

The tanh function is given by 
\begin{equation*} 
    \mathrm{tanh}(x) = \phi(x) = \frac{e^{x} - e^{-x}}{e^{x} + e^{-x}}.
\end{equation*}
The process of computing the sectors is the same as the sigmoid function, however they are positioned differently. As in the sigmoid case, the process for computing these bounds is presented in \cite{mnew2} and shown visually in Figure \ref{fig:tanhsec}. 

\section{Problem Formulation} \label{sec:problem}
As alluded to previously, most recent works consider a particular type of optimisation framework to verify the neural network. However, when neural networks have varying sizes with different activation functions a more general approach is needed. This will not only make it easier to express and implement the optimisation problem for an arbitrary neural network architecture, but it also allow us to more freely trade off accuracy with scalability. To achieve this we use a theorem from real algebraic geometry known as the Positivstellensatz (Psatz) \cite{gsten}, that uses an algebraic condition to test the emptiness of a semi-algebraic set. We now describe this theorem and how it can be applied to the neural network verification problem.
\subsection{Positivstellensatz}
The Positivstellensatz provides a link between the emptiness of a semi-algebraic set and an algebraic condition. A basic closed semi-algebraic set is defined by 
\begin{equation*}
    \big\{ x\in \mathbb{R}^{n} \: |\:  f_{i}(x) \geq 0 \: \: \forall \: i = 1, \dots, m \big\}.
\end{equation*}
where $f_i(x)$ are polynomials in $x$, i.e. $f_i(x) \in \mathbb{R}[x]$. This can be extended to define the set 
\begin{equation*}
   S = \big\{ x \in \mathbb{R}^{n} \: |\:  f_{i}(x) \star 0 \: \: \forall \: i = 1, \dots, m \big\},
\end{equation*}
where $\star$ denotes $<,\leq, =,$ or $ \neq $. In this paper we use the following notation to describe a semi-algebraic set: 
\begin{equation} 
    S = \big\{ x \in \mathbb{R}^{n} \: | \:  g_{i}(x) \geq 0, \: h_{j}(x) = 0, \: \: \forall \: i = 1, \dots, p, \: j = 1, \dots , q \big\}, \label{Sset}
\end{equation}
where $g_i(x), h_j(x) \in \mathbb{R}[x]$.
\begin{definition} \label{def:sos}
    A polynomial $p(x)$ is said to be a sum of squares (SOS) polynomial if it can be expressed as 
    \begin{equation*}
        p(x) = \sum_{i=1}r_{i}^{2}(x) \equiv p(x) \: \mathrm{is} \: \mathrm{SOS}.
    \end{equation*}
    We denote the set of polynomials that admit this decomposition by $\Sigma[x]$.
\end{definition}
\begin{definition} \label{def:cone}
    The \emph{cone} of a set of polynomials is defined as
    \begin{equation*}
        \mathrm{cone}\{g_{1}, \dots , g_{p} \} = \Bigg\{ \sum_{i=1}^{p} s_{i}g_{i} \: | \:  s_{i} \in \Sigma [x],  g_{i} \in \mathbb{R}[x] \Bigg\}.
    \end{equation*}
\end{definition}
\begin{definition} \label{def:ideal}
    The \emph{ideal} of a set of polynomials is defined as 
    \begin{equation*}
        \mathrm{ideal}\{h_{1}, \dots , h_{q} \} = \Bigg\{ \sum_{j=1}^{q} t_{j}h_{j} \: | \: t_{j} \in \mathbb{R}[x] \Bigg\}.
    \end{equation*}
\end{definition}
\begin{theorem} (Positivstellensatz) \label{psatz1}
    Given the semi-algebraic set $S$ defined in \eqref{Sset}, the following are equivalent:
    \begin{enumerate}
        \item The set $S$ is empty.
        \item There exist $s_i \in \Sigma[x]$ and $t_j \in \mathbb{R}[x]$ such that $-1 \in \mathrm{cone}\{g_{1}, \dots , g_{p} \} + \mathrm{ideal} \{h_{1}, \dots , h_{q} \}$. 
    \end{enumerate}
\end{theorem}
Theorem \ref{psatz1} links the emptiness of a semi-algebraic set with an algebraic test. There are many different formulations of this theorem, one way is to attempt a representation of the function $f$ such that if
\begin{equation}
    f = 1 + \sum_{j}^{q}t_{j}h_{j} + s_{0} + \sum_{i}^{p}s_{i}g_{i} + \\ \sum_{i\neq j}^{p}r_{ij}g_{i}g_{j} + \sum_{i\neq j \neq k}^{p}r_{ijk}g_{i}g_{j}g_{k} + \dots \label{eq:psatz}
\end{equation}
then $f(x)>0, \: \forall \: x \in S$, where $s_{i},r_{ij},r_{ijk} \dots \in \Sigma[x]$ and $t_{j}\in \mathbb{R}[x]$.

\subsection{Neural Network Verification Emptiness Condition}
To set up the neural network verification problem we adjust the Psatz condition slightly to use it more easily in conjunction with the optimisation framework. Instead of showing that $g_{out}^{m} \geq 0$ in a feasibility test, using the Psatz we show that $g_{out}^{m} < 0$ is infeasible. We set $\gamma_{m}$ as the decision variable in the optimisation program so it can be optimised to find the limiting value to when this emptiness condition is violated. We can then write the Psatz conditions as:
\begin{equation*}
\begin{aligned}
    \mathrm{minimize} \quad & \gamma_{m}, \\
    \mathrm{subject \: to} \quad & -c_{m}^{T}y + \gamma_{m} - \sum_{j}^{q}t_{j}h_{j} - \sum_{i}^{p}s_{i}g_{i} - \sum_{i\neq j}^{p}r_{ij}g_{i}g_{j} - \dots \: \mathrm{is \: SOS}, \\
    \quad & s_{i} \: \mathrm{is \: SOS},\: \forall \: i = 1,\dots, p, \: \: \: \: \: \: r_{ij} \: \mathrm{is \: SOS},\: \forall \: i,j = 1,\dots, p, 
    \: \: \: \: \: \: t_{j} \in \mathbb{R}[x], \:  \forall \: j = 1,\dots, q,
\end{aligned}
\end{equation*}
where $h_{j}$ and $g_{i}$ are the equality and inequality constraints respectively.

To test the emptiness of semi-algebraic sets through the Psatz computationally, one can use polynomial optimisation and SOS to check the algebraic condition - we describe this process in more detail in the next section. The optimisation problem results into a set of SOS conditions, which can be checked using SOSTOOLS \cite{sostools} in MATLAB or the SumOfSquares.jl package \cite{sosjl} in Julia. If we choose a higher degree for the multipliers $s_i$, $t_j$ etc., we can obtain a series of nested set emptiness tests of increasing complexity and non-decreasing accuracy. 

\subsection{Sum of Squares}
SOS conditions are useful since they can be cast into Linear Matrix Inequality (LMI) constraints and then solved using semidefinite programming (SDP) \cite{ppar}. Instead of checking the nonnegativity of a polynomial which is known to be an NP-hard problem \cite{kmur}, we can check if a polynomial is SOS, which can be done by solving an equivalent semidefinite program (SDP) in polynomial time. This is achieved by creating a monomial vector which contains a selection of the variables $x=[x_{1}, \dots x_{n}]$. A monomial defined by all $n$ variables is denoted as $x^{\beta} = x_{1}^{\beta_{1}}x_{2}^{\beta_{2}} \dots x_{n}^{\beta_{n}}$, where the exponent and degree are denoted as $\beta = (\beta_{1}, \dots , \beta_{n}) \in \mathbb{N}^{n}$ and $|\beta| = \beta_{1} + \dots + \beta_{n}$ respectively; $\mathbb{N}^{n}$ denotes the set of $n$ integers. We express the column vector of monomials with only certain exponents as $x^{\mathbb{B}} = (x^{\beta})_{\beta \in \mathbb{B}}$, where $\mathbb{B} \subset \mathbb{N}^{n}$ is the set of exponents that are used in the monomials. Note that any polynomial $f$ can be written as $f = \sum_{\beta \in \mathbb{N}_{d}^{n}}f_{\beta} x^{\beta}$ for a set of coefficients $f_{\beta} \in \mathbb{R}$, where $\mathbb{N}_{d}^{n} = \{ \beta \in \mathbb{N}^{n} \: : \: |\beta| \leq d \}$ is the set of all $n$-variate exponents of degree $d$ or less. 
We also define the summation operation on $\mathbb{B}$ as
\begin{equation*}
    \mathbb{B} + \mathbb{B} := \{ \beta + \gamma \: : \: \beta, \gamma \in \mathbb{B} \}.
\end{equation*}
The sets of symmetric and positive semidefinite matrices are denoted by $\mathbb{S}^{n}$ and $\mathbb{S}_{+}^{n}$ respectively. A polynomial $f$ is SOS if and only if it can be written in what is referred to as a Gram matrix representation such that $f = (x^{\mathbb{B}})^{T} Q x^{\mathbb{B}}$, where  $Q \in \mathbb{S}_{+}^{|\mathbb{B}|}$ is a positive semidefinite matrix. The existence of an SOS decomposition for a polynomial is only a sufficient condition for global non-negativity: the Motzkin polynomial is a well known example of a nonnegative polynomial that is not representable as a SOS \cite{tmot}. To convert the Gram matrix representation into SDP constraints we first define the symmetric binary matrix $A_{\alpha} \in \mathbb{S}^{|\mathbb{B}|}$ for each exponent $\alpha \in \mathbb{B} + \mathbb{B}$ as
\begin{equation*}
    [A_{\alpha}]_{\beta,\gamma} := \begin{cases}
    1, \: \beta + \gamma = \alpha, \\
    0, \: \mathrm{otherwise}.
    \end{cases}
\end{equation*}
We can rewrite the Gram representation as
\begin{equation*}
    (x^{\mathbb{B}})^{T} Q x^{\mathbb{B}} = \langle Q, x^{\mathbb{B}} (x^{\mathbb{B}})^{T} \rangle = \sum_{\alpha = \mathbb{B} + \mathbb{B}} \langle Q, A_{\alpha} \rangle x^{\alpha}.
\end{equation*}
Therefore, the following is true
\begin{equation*}
    f \in \Sigma[x] \Leftrightarrow 
    \exists Q \in \mathbb{S}_{+}^{|\mathbb{B}|} \: \mathrm{such \: that} \:  \langle Q, A_{\alpha} \rangle =f_{\alpha}, \: \forall \alpha \in \mathbb{B} + \mathbb{B}.
\end{equation*}

In the default case the vector of monomials $x^{\mathbb{B}}$ contains all monomials of degree up to $\mathrm{deg}(\frac{1}{2}f)$ and the matrix $Q$ is a fully dense matrix of size ${n+d \choose d}  \times {n+d \choose d}$. However, in a system that possesses a significant level of sparsity then it is expected that not all of the monomial terms may be needed. Hence if only a subset of them are included, the size of the resulting SDP will be reduced; this idea is explored in more detail in Section \ref{sec:sparsePOP}.

\section{Sparse Polynomial Optimisation} \label{sec:sparsePOP}

\subsection{Overview}
One of the issues with using an SDP framework is that the computational time becomes too large as the size of the neural network gets bigger: this is what we refer to as the scalability issue. Since neural networks in practice have shown great success when the number of nodes in each layer is large and when there are numerous layers, it is important to devise ways to verify their properties using semidefinite programming that also have tractable solve times. 

\begin{figure}[h]
    \centering 
    \includegraphics[height=7.5cm]{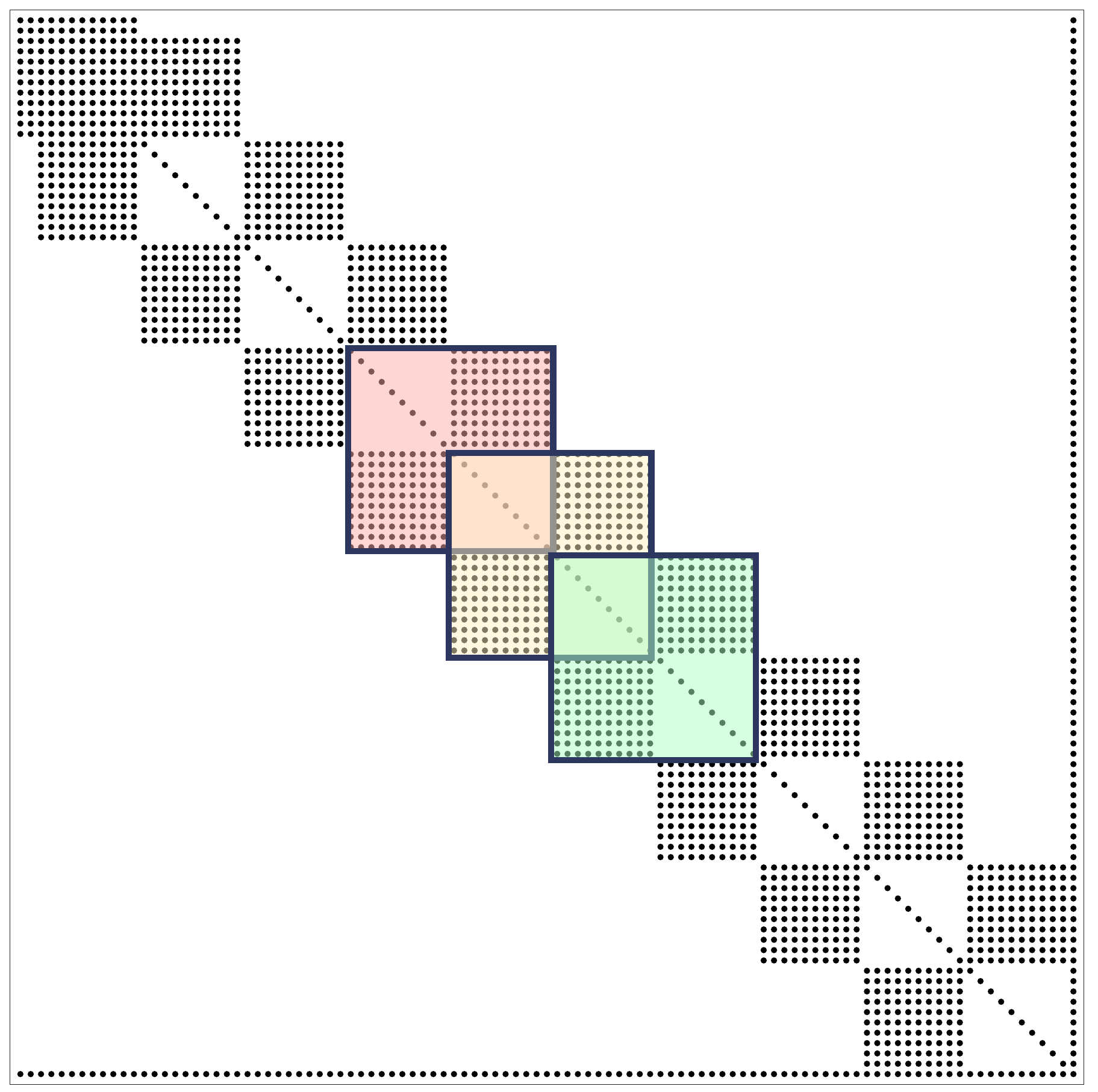}
    \caption{Sparsity pattern of the positive semidefinite constraint matrix of a ten layer neural network with ten nodes in each layer. The blocks show the constraints from overlapping layers.} \label{fig:matrix}
\end{figure}

We first note that a fully connected feed-forward neural network possesses a natural cascading structure, where each layer is only connected directly to adjacent layers. In fact, common constraints that are used to bound the activation functions only contain variables in the current layer and the previous layer. This means that the constraint matrices have a very well-defined structure and therefore this cascading structure is inherited in the algorithm formulation. For example, consider a neural network with two inputs, two outputs, ten nodes in each layer each with ten hidden units. If we use quadratic bounds for the ReLU activation function and the S-procedure as in \cite{mfaz} (which is a specific case of the Psatz) \cite{vyak}, the constraint matrix in the SDP contains a significant amount of sparsity, as most of the terms are zero. This is shown in Figure \ref{fig:matrix} \cite{mnew}, and it can be exploited computationally to greatly improve the solve time of the SDP, as we show next.

\subsection{Chordal Graphs and Sparse Matrix Decomposition} \label{sec:4}
We now provide an overview of the theory of chordal graphs and the ways in which they can be used to exploit the sparsity in positive semidefinite matrices, and therefore how they can be used to improve the solve time of an SDP. We will show that the neural network verification problem possesses this chordal sparsity property, motivating the use of sparse matrix decomposition to overcome scalability issues.

A graph $\mathcal{G}(\mathcal{V},\mathcal{E})$ is defined as a set of vertices $\mathcal{V} = \{1,2,\dots,n \} $ and a set of edges $\mathcal{E} \subseteq	\mathcal{V} \times \mathcal{V} $. A vertex-induced subgraph $G^{'}(\mathcal{V}^{'},\mathcal{E}^{'})$ is a graph with a subset of the vertices of the graph $G(\mathcal{V},\mathcal{E})$ together with any edges whose endpoints are both in this subset. A clique $\mathcal{C} \subseteq \mathcal{V}$ is a subgraph such that all the vertices in the subgraph $\mathcal{C}$ form a complete graph - a complete graph is a graph such that any two nodes are connected by an edge. A maximal clique is a clique that is not a subset of any other clique. A graph can contain a cycle, which is defined by a set of pairwise distinct nodes $\{ v_{1}, v_{2}, \dots, v_{k} \} \subset \mathcal{V} $ such that $ (v_{k}, v_{1}) \in \mathcal{E} $ and $(v_{i}, v_{i+1}) \in \mathcal{E}$ for $i=1,\dots,k-1$. A chord that lies on the graph $\mathcal{G}(\mathcal{V},\mathcal{E})$ is an edge that joins two non-adjacent nodes in a cycle \cite{nkak}.

\begin{definition} A connected undirected graph $\mathcal{G}(\mathcal{V},\mathcal{E})$ is \emph{chordal} if every cycle of length four or greater has at least one chord.
\end{definition}

Chordal graphs are useful since they can be decomposed into their maximal cliques \cite{lvan}. A graph that is not chordal can be extended to become chordal by adding additional edges to take advantage of this well defined structure:

\begin{definition} 
The \emph{chordal extension} of a graph $\mathcal{G}(\mathcal{V},\mathcal{E})$ is denoted as $\hat{\mathcal{G}}(\mathcal{V},\hat{\mathcal{E}})$, where $\mathcal{E} \subseteq \hat{\mathcal{E}} $ and $\hat{\mathcal{G}}$ is chordal. 
\end{definition} 

Consider now a symmetric matrix $X \in \mathbb{S}^{n}$ with a sparsity pattern represented by an undirected graph $\mathcal{G}(\mathcal{V},\mathcal{E})$, such that $ X_{ij} = X_{ji} = 0, \: \forall i \neq j$  if $ (i,j) \notin \mathcal{E}$. This means that the matrix $X$ has a zero entry in elements that correspond to the nodes that are not connected by edges on the graph. Just as a chordal graph can be decomposed into its maximal cliques, a matrix $X$ with a chordal sparsity pattern can be split up into smaller sub-matrices, with the sub-matrices corresponding to the maximal cliques of the chordal graph. 
An important result relates matrices $X$ that are positive semidefinite, to such a decomposition. 

\begin{figure}[h]
    \centering 
    \includegraphics[height=4cm]{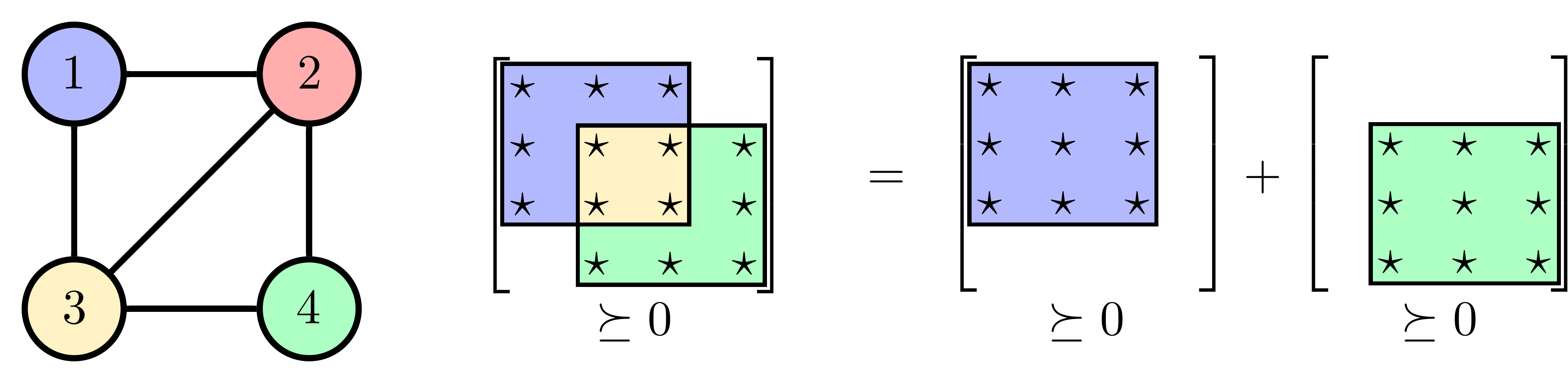}
    \caption{Shows how a matrix can be represented by a graph and then how a positive semidefinite constraint can be split into smaller positive semidefinite constraints using the properties of the chordal graph.} \label{fig:chordalgraph}
\end{figure}

\begin{theorem} \cite{agler}  Consider the chordal graph $\mathcal{G}(\mathcal{V},\mathcal{E})$ that is made up of maximal cliques $ \{ \mathcal{C}_{1}, \mathcal{C}_{2}, \dots, \mathcal{C}_{t} \}$. Then $ Z \in \mathbb{S}^{n}_{+}(\mathcal{E},0)$ if and only if there exist $Z_{k} \in \mathbb{S}^{|\mathcal{C}_{k}|}_{+}$ for $k = 1,\dots,t$ such that \label{theorem:alger}
\begin{equation*}
    Z = \sum_{k=1}^{t} E_{\mathcal{C}_{k}}^{T} Z_{k} E_{\mathcal{C}_{k}},
\end{equation*}
where $\mathbb{S}^{n}_{+}(\mathcal{E},0) := \{ X \in \mathbb{S}^{n}\: |\: X \succeq 0 \: | \: X_{ij} = X_{ji} = 0,\: \mathrm{if} \: i \neq j \: \mathrm{and} \: (i,j) \notin \mathcal{E} \} $, $|\mathcal{C}_{i}|$ is the number of vertices in that clique and 
\begin{equation*}
    \left( E_{\mathcal{C}_{k}} \right)_{ij} = 
    \begin{cases} 1, & \mathrm{if} \: \mathcal{C}_{k}(i)=j \\ 0, & \mathrm{otherwise.}
    \end{cases}
\end{equation*}
\end{theorem}
This is useful as it means we can test that a large matrix with chordal sparsity is positive semidefinite in a distributed way. Theorem \ref{theorem:alger} is shown visually in Figure \ref{fig:chordalgraph}. This idea can be further extended to sparse block matrices \cite{yzhe2}. 

The chordal sparsity property can be used to replace a large positive semidefinite condition of a single large matrix in an SDP with multiple positive semidefinite conditions of smaller size. There are different solvers that exist to achieve this, one example is sparseCoLo \cite{sparsecolo}, which incorporates four conversion methods. It can be used with both the primal and dual forms of linear, semidefinite and second-order cone programs that have both equality and inequality constraints. CDCS \cite{cdcs} is another solver that uses a first order splitting method called alternating direction method of multipliers (ADMM). This solver scales better to large systems but often provides less accurate solutions. Another solver that uses an operator splitting method is the Conic Operator Splitting Method (COSMO) \cite{cosmo} for convex optimisation problems with a quadratic objective function and conic constraints. COSMO uses chordal decomposition and a clique merging algorithm to exploit the sparsity of the problem. Other examples of solvers include SMCP \cite{smcp} and SDPA-C \cite{SDPAC}. 

\subsection{Term Sparsity}
For a more comprehensive review of term sparsity, the reader is referred  to \cite{yzhe3, jwan, jwan2}. Term sparsity involves reducing the size of the support set of $f$, which is defined as
\begin{equation*}
    \mathrm{supp}(f) = \{ \beta \in \mathbb{N}_{d}^{n} \: : \: f_{\beta} \neq 0 \}.
\end{equation*}
The simplest way to exploit term sparsity is to take the Newton polytope reduction such that 
\begin{equation*}
    \mathbb{B} = \frac{1}{2} \mathrm{New}(f) \cap \mathbb{N}_{d}^{n}. 
\end{equation*}
where $\mathrm{New}(f)$ of $f$ is the convex hull of $\mathrm{supp}(f)$. This can be simplified further using general facial reduction techniques such as \cite{jlof} and \cite{fper}. Such techniques will remove redundant elements of $\mathbb{B}$ to construct a smaller exponent set. However, facial reduction can sometimes only reduce a very small number of terms which might still mean that some problems remain intractable. More advanced techniques exist to reduce this term sparsity further, and \cite{yzhe3} describes a general approach to exploiting term sparsity. An important point is that although these sparse representations will reduce the computational complexity, some are conservative, introducing a trade-off, as term sparsity in the reduced support set does not imply the existence of an SOS decomposition.

Consider a graph $\mathcal{G}(\mathbb{B}, \mathcal{E})$ with maximal cliques $\mathcal{C}_{1}, \dots, \mathcal{C}_{t}$ and edge set $\mathcal{E} \subseteq \mathbb{B} \times \mathbb{B}$. The exponent set can be written as 
\begin{equation*}
    \mathbb{A} \subseteq \{ \beta + \gamma \: : \: (\beta, \gamma) \in \mathcal{E} \}.
\end{equation*}
Given this exponent set, define the subcone of SOS polynomials that are supported on $\mathbb{A}$ as
\begin{equation*}
     \Sigma [\mathbb{A}] := \{ f \in \Sigma \: : \: \mathrm{supp}(f) \subseteq \mathbb{A} \}.
\end{equation*}
We can then use the clique-based positive semidefinite decomposition to express the Gram matrix as 
\begin{equation*}
    Q = \sum_{k=1}^{t} E_{\mathcal{C}_{k}}^{T}Z_{k}E_{\mathcal{C}_{k}}, \: \mathrm{where } \: Z_{k} \in \mathcal{S}_{+}^{|\mathcal{C}_{k}|}.
\end{equation*}
This $Q$ belongs to the cone of sparse SOS polynomials expressed as
\begin{equation*}
    \Sigma [\mathbb{A};\mathcal{E}] := \{ f \in \Sigma [\mathbb{A}] \: : \: f = (x^{\mathbb{B}})^{T} Q x^{\mathbb{B}} \},
\end{equation*}
and it can be shown that $\Sigma[\mathbb{A};\mathcal{E}] \subseteq \Sigma[\mathbb{A}]$. The cone can be converted into an SDP condition such that $f \in \Sigma[\mathbb{A};\mathcal{E}]$ if and only if
\begin{equation*}
    \exists Z_{1} \in \mathbb{S}_{+}^{|\mathcal{C}_{1}|}, \dots , Z_{t} \in \mathbb{S}_{+}^{|\mathcal{C}_{t}|}, \: \mathrm{such \: that} \: \sum_{k=1}^{t} \langle Z_{k}, E_{\mathcal{C}_{k}} A_{\alpha} E_{\mathcal{C}_{k}}^{T} \rangle = f_{\alpha} \quad \forall \alpha \in \mathbb{B} + \mathbb{B}.
\end{equation*}
The difficulty is how to select the cliques in the correct way for this decomposition to be valid. For non-chordal graphs this is an NP-hard problem, however chordal extensions can be created to overcome this. In our work however, we know \emph{a priori} the structure of the problem and therefore we can directly analyse and select the cliques, so that we know they are chordal. However, term sparsity can be restrictive as it may not include enough terms for the SOS decomposition to exist; less conservative formulations can be considered. 

\subsection{Correlative Sparsity}
Correlative sparsity takes a different approach to term sparsity. Instead of trying to reduce the size of $|\mathrm{supp}(f)|$, it considers couplings between variables. Two variables are considered to be coupled if a monomial term depends on both variables simultaneously. The correlative sparsity graph of the support set $\mathbb{A}$ is defined by
\begin{equation*}
    \mathcal{S}_{csp}(\mathbb{A}) := \{ (i,j)\: : \: \exists \alpha \in \mathbb{A} \: \mathrm{with} \: \alpha_{i}\alpha_{j} > 0 \},
\end{equation*}
i.e. the two variables $x_{i}$ and $x_{j}$ corresponding to $\alpha_{i}$ and $\alpha_{j}$ respectively are considered to be coupled. When applied to the sparse SOS decomposition, the entries in the $Q$ matrix that correspond to couplings that do not belong to the correlative sparsity graph of $f$ are set to zero. The sparsity graph of $Q$ then becomes $\mathcal{G}_{csp}(\mathbb{B}, \mathcal{E}_{csp})$ where the edge set is defined by
\begin{equation*}
    \mathcal{E}_{csp} := \{ (\beta, \gamma) \in \mathbb{B} \times \mathbb{B}\: : \: (\beta_{i} + \gamma_{i})(\beta_{j} + \gamma_{j}) >0 \: \Rightarrow (i,j) \in \mathcal{S}_{csp}(\mathbb{A}) \}.
\end{equation*}
From $\mathcal{G}_{csp}(\mathbb{B}, \mathcal{E}_{csp})$ we can build $\mathcal{G}(\mathbb{B}, \mathcal{E}_{csp})$ ensuring that polynomials $(x^{\mathbb{B}})^{T} Q x^{\mathbb{B}}$ with $Q \in \mathbb{S}^{|\mathbb{B}(\mathcal{E}_{csp},0)|}$ inherit the correlative sparsity of the original support set $\mathbb{A}$. It can be shown that the properties of $\mathcal{G}(\mathbb{B}, \mathcal{E}_{csp})$ can be inferred from $\mathcal{G}_{csp}(\mathbb{B}, \mathcal{E}_{csp})$ \cite{jwan2}.
\begin{theorem} \cite{yzhe3}
Consider a correlative sparsity graph that has a support set $\mathbb{A}$ and maximal cliques $\mathcal{J}_{1}, \dots \mathcal{J}_{t}$, then $\mathcal{G}(\mathbb{B}, \mathcal{E}_{csp})$ has maximal cliques $\mathcal{C}_{k} = \{ \beta \in \mathbb{B} \: : \: \mathrm{nnz}(\beta) \subseteq \mathcal{J}_{k} \}$ for $k=1, \dots , t$, where $\mathrm{nnz}(\beta) := \{ \beta \in \mathbb{B} : \beta_{i} \neq 0, \: \forall i = 1, \dots , |\mathbb{B}| \}$. Moreover, if the correlative sparsity graph of $\mathbb{A}$ is chordal then so is $\mathcal{G}(\mathbb{B}, \mathcal{E}_{csp})$. 
\end{theorem}
This is useful as typically $\mathcal{G}_{csp}(\mathbb{B}, \mathcal{E}_{csp})$ is smaller than $\mathcal{G}(\mathbb{B}, \mathcal{E}_{csp})$, which makes finding the maximal cliques easier. These cliques can then be converted into LMI constraints to be solved in an SDP as described previously.

\subsection{Similar Hierarchies}
Correlative sparsity can sometimes not account for the full structure of $\mathbb{A}$, especially when a significant amount of term sparsity exists. To overcome this, different hierarchies to define the monomial basis have been proposed. Most notable ones are Term Sparse SOS (TSSOS), Chordal-TSSOS (CTSSOS) and Correlative Sparsity-TSSOS (CS-TSSOS), which can exploit term sparsity even when $f$ is not necessarily correlatively sparse \cite{jwan, jwan2}. The general techniques of these approaches is to iteratively update the graph to form the hierarchy, starting from the fact that each edge set should contain at least all edges $(\beta, \gamma)$ with $\beta + \gamma \in \mathbb{A}$ since it guarantees that $\mathbb{A} \subseteq \mathrm{supp} ((x^{\mathbb{B}})^{T} Q x^{\mathbb{B}})$. For details of this iterative scheme the readers should refer to \cite{jwan} and \cite{jwan2}. 

\subsection{Extension to Semi-algebraic Sets}
The above methods to construct SOS decompositions by exploiting sparsity only show global nonnegativity, however in the neural network verification problem we require this theory to be applied to semi-algebraic sets and establish local nonnegativity. In the general case we can define the semi-algebaric set with $m$ polynomial inequalities such that
\begin{equation*}
    S := \{ x \in \mathbb{R}^{n} \: : \: g_{1}(x) \geq 0, \dots , g_{m}(x) \geq 0  \}.
\end{equation*}
Using the Psatz as in \eqref{eq:psatz} it is possible to verify that $f \in \mathbb{R}[x]$ is nonnegative on $S$. Consider a series of exponent sets associated with each inequality constraint $\mathbb{B}_{0}, \dots , \mathbb{B}_{m} \subseteq \mathbb{N}_{\omega}^{n}$ where $\omega$ is known as the relaxation order. The Psatz for this case with no equality constraints and where no inequality constraints are multiplied together states that if 
\begin{equation} \label{sparsePsatz}
    f(x) = \sum_{i=0}^{m} g_{i}(x) (x^{\mathbb{B}_{i}})^{T} Q_{i} x^{\mathbb{B}_{i}}, \quad Q_{i} \in \mathbb{S}_{+}^{|\mathbb{B}_{i}|}.
\end{equation}
then $f(x)$ is non-negative. The relaxation order $\omega$ is a parameter to be chosen; as its value increases the solutions do not become less accurate and in many cases become better; however this will also increase the size of the SDP. One must be careful not to make $\omega$ too large otherwise the problem may become computationally intractable, but not too small to sacrifice solution accuracy significantly. A common choice for $\omega$ is 
\begin{equation*}
    2 \omega \geq \mathrm{max} \{ \mathrm{deg}(f), \mathrm{deg}(g_{1}), \dots , \mathrm{deg}(g_{m})  \},
\end{equation*}
such that the exponent set becomes
\begin{equation*}
    \mathbb{B}_{i} = \mathbb{N}_{\omega_{i}}^{n}, \: \mathrm{where} \: \omega_{i} := \omega - \lceil \frac{1}{2}\mathrm{deg}(g_{i}) \rceil.
\end{equation*}
For global nonnegativity we only need to consider the sparsity graph of $f$, however the challenge in the case of positivity over $S$ is that we must consider how the sparse polynomial multiplier $(x^{\mathbb{B}_{i}})^{T} Q_{i} x^{\mathbb{B}_{i}}$ interacts with the corresponding inequality constraint $g_{i}(x)$. If the $\mathbb{B}_{i}$ destroys the sparsity in \eqref{sparsePsatz} then we cannot exploit the sparsity in $f(x)$ or the $g_{i}(x)$ inequality constraints. 

To proceed to preserve and analyse the sparsity pattern, we replace the correlative sparsity graph \cite{yzhe3} of $f$ with a joint correlative sparsity graph of the polynomials $f, g_{1}, \dots , g_{m}$. As before this graph has $n$ vertices, however the edges are constructed differently. An edge between vertices $i$ and $j$ exists if at least one of the following are true:
\begin{itemize}
    \item[] Condition 1. The variables $x_{i}$ and $x_{j}$ are multiplied together in $f$.
    \item[] Condition 2. At least one of the $g_{1}, \dots , g_{m}$ depends on both $x_{i}$ and $x_{j}$, even if these variables are not multiplied together.
\end{itemize}
Therefore, the support of $g_{i}(x)(x^{\mathbb{B}})^{T} Q_{i} x^{\mathbb{B}}$ must be consistent with the joint correlative sparsity graph, where each matrix $Q_{i}$ is the densest possible matrix. The maximal cliques of the joint correlative sparsity graph are defined to be $\mathcal{J}_{1}, \dots , \mathcal{J}_{t}$. Through Condition 1, we can be sure that there is at least one clique $\mathcal{J}_{k}$ such that  $\mathrm{var}(g_{i}) \subseteq \mathcal{J}_{k}$, where  $\mathrm{var}(g_{i}) \subset \{ 1, \dots , n \}$ is the set of indices of the variables on which $g_{i}$ depends. The set of cliques for which this holds is denoted by 
\begin{equation*}
    \mathcal{N}_{i} := \{ k \in \{ 1, \dots , t \} \: : \: \mathrm{var}(g_{i}) \subseteq \mathcal{J}_{k} \}.
\end{equation*}
The edges of the sparsity graph $\mathcal{G}_{i}(\mathbb{B}_{i}, \mathcal{E}_{i})$ corresponding to $Q_{i}$ are defined as 
\begin{equation*}
    \mathcal{E}_{i} := \bigcup_{k \in \mathcal{N}_{i}} \{ (\beta, \gamma) \in \mathbb{B}_{i} \times \mathbb{B}_{i} \: : \: \mathrm{nnz}(\beta + \gamma) \subseteq \mathcal{J}_{k} \}.
\end{equation*}
If this graph is chordal then it has maximal cliques $\mathcal{C}_{i,1}, \dots ,\mathcal{C}_{i,|\mathcal{N}_{i}|} $ where $\mathcal{C}_{i,k} := \{ \beta \in \mathbb{B}_{i} \: : \: \mathrm{nnz}(\beta) \subseteq \mathcal{J}_{k} \}$. Hence the positive semidefinite decomposition can be written as
\begin{equation*}
    Q_{i} = \sum_{k=1}^{|\mathcal{N}_{i}|} E_{\mathcal{C}_{i,k}}^{T} Z_{k} E_{\mathcal{C}_{i,k}}, \quad Z_{k} \in \mathbb{S}_{+}^{|\mathcal{C}_{i,k}|}.
\end{equation*}
The following procedure can be extended to TSSOS and CS-TSSOS hierarchies and the result is similar in that the local formulation stabilizes to a particular hierarchy in the same way as in the global scheme.  The details of these processes are more involved and will be omitted for conciseness, but more details can be found in \cite{jwan} and \cite{jwan2}.

\section{Sparse Neural Network Constraints} \label{sec:sparseNN}
We now use the theory outlined in Section \ref{sec:sparsePOP} in conjunction with the Psatz condition in Section \ref{sec:problem} to formulate a sparse version of the neural network verification problem. Since the neural network has a natural cascading structure, it is possible to construct the semi-algebraic constraints to be only a function of a single layer and the previous layer. We show how this sparsity arises with a simple example. 

\subsection{Example: Three Layer, Single Node Neural Network} \label{subsec:NNexample}
\begin{figure}[h] 
    \centering
    \includegraphics[height=4.5cm]{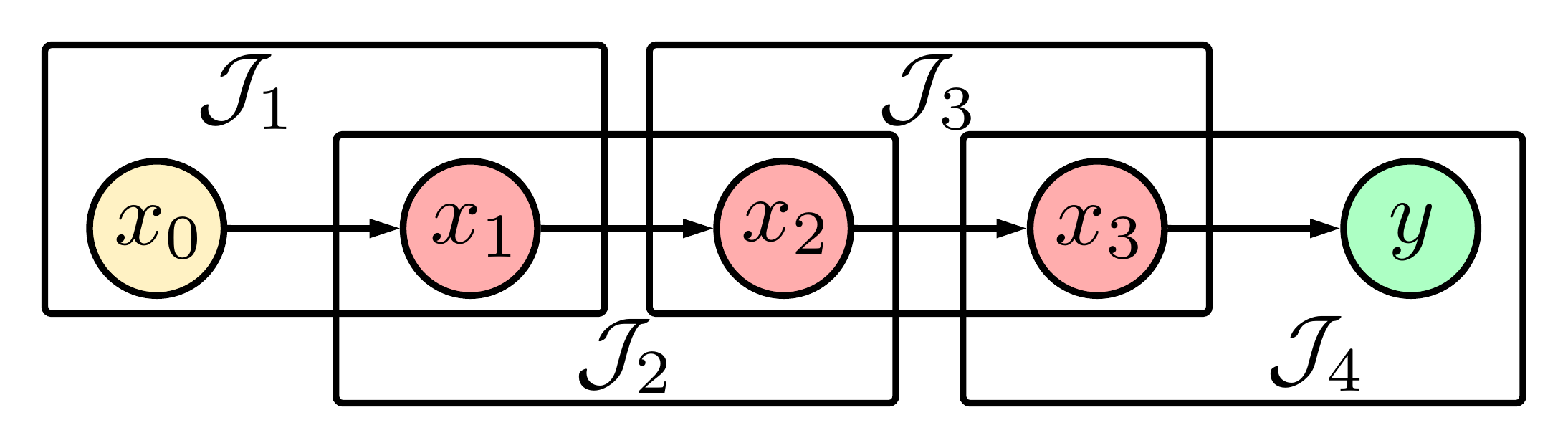}
    \caption{Diagram of a neural network showing how the cliques are formed for the example in Section \ref{subsec:NNexample}.} \label{fig:NNexample}
\end{figure}

Consider a single input/output neural network with three layers and a single node in each layer with a ReLU activation function. The equations for this neural network are:
\begin{equation*}
    x_{0} = u, \: x_{1} = \phi (W^{0}x_{0} + b^{0}), \: x_{2} = \phi (W^{1}x_{1} + b^{1}), \: x_{3} = \phi (W^{2}x_{2} + b^{2}), \: y = W^{3}x_{3} + b^{3}.
\end{equation*}
For the constraints we use the notation $g_{i,j}$ and $h_{i,j}$ to represent the $j^{th}$ inequality and equality constraint respectively in the $i^{th}$ layer. We set the input to be bounded by $[-1,1]$, therefore the input constraints are 
\begin{equation*}
    g_{0,1}(x_{0}) = x_{0} + 1 \geq 0, \: g_{0,2}(x_{0}) = 1 - x_{0} \geq 0.
\end{equation*}
For now we will only consider the ReLU activation function in the hidden layers. We will use the quadratic constraints as outlined in \eqref{eq:relu} such that
\begin{equation*}
    x_{i} \geq 0, \: x_{i} - (W^{i-1}x_{i-1} + b^{i-1}) \geq 0, \: x_{i}(x_{i} - (W^{i-1}x_{i-1} + b^{i-1})) = 0, \: \mathrm{for} \: i = 1,2,3.
\end{equation*}
The constraints for this example are therefore:
\begin{equation*}
     g_{0,1}(x_{0}) = x_{0} + 1 \geq 0, \: g_{0,2}(x_{0}) = 1 - x_{0} \geq 0,
\end{equation*}
\begin{equation*}
    g_{1,1}(x_{1}) = x_{1} \geq 0, \: \: g_{1,2}(x_{1},x_{0}) = x_{1} - (W^{0}x_{0} + b^{0}) \geq 0, \: \: h_{1,1}(x_{1},x_{0}) = x_{1}^{2} - x_{1}(W^{0}x_{0} + b^{0}) = 0, 
\end{equation*}
\begin{equation*}
    g_{2,1}(x_{2}) = x_{2} \geq 0, \: \: g_{2,2}(x_{2},x_{1}) = x_{2} - (W^{1}x_{1} + b^{1}) \geq 0, \: \: h_{2,1}(x_{2},x_{1}) = x_{2}^{2} - x_{2}(W^{1}x_{1} + b^{1}) = 0,
\end{equation*}
\begin{equation*}
    g_{3,1}(x_{3}) = x_{3} \geq 0, \: \: g_{3,2}(x_{3},x_{2}) = x_{3} - (W^{2}x_{2} + b^{2}) \geq 0, \: \: h_{3,1}(x_{3},x_{2}) = x_{3}^{2} - x_{3}(W^{2}x_{2} + b^{2}) = 0,
\end{equation*}
\begin{equation*}
    g_{4,1}(x_{4},x_{3}) = y - (W^{3}x_{3} + b^{3}) \geq 0.
\end{equation*}
If we start with the simplest case and choose the relaxation order to be zero ($\omega = 0$), then the monomial basis for the multipliers will always be unity ($x^{\mathbb{B}_{i}} = 1$). We can then form a graph with the variables that are connected to one another. If we consider the case of correlative sparsity then the graph becomes a line graph as shown in Figure \ref{fig:NNexample}. This graph has four cliques, which are written as:
\begin{equation*}
    \mathcal{J}_{1} = \{0, 1 \}, \: \mathcal{J}_{2} = \{1, 2 \}, \: \mathcal{J}_{3} = \{2, 3 \}, \: \mathcal{J}_{4} = \{3, 4 \}.
\end{equation*}

We now consider what happens when we increase the relaxation order and how we can select the correct exponents. Condition 1 states that an edge between vertices $i$ and $j$ is connected if the variables $x_{i}$ and $x_{j}$ are multiplied together in $f$; one can see easily that due to the structure of $f$ this condition will never be true. Therefore, our attention turns to the second condition that states that a connection exists if at least one of $g_{i,j}, \: \forall i,j$ depends on both $x_{i}$ and $x_{j}$ even if these variables are not multiplied together. So for the connection between $x_{0}$ and $x_{1}$, the only constraints that depend on these two variables are the constraints in the first layer, i.e., $g_{1,2}$ and $h_{1,1}$. We now impose that the $Q_{i,j}$ matrix is as dense as possible such that the support of $g_{i,j}(x)(x^{\mathbb{B}_{i,j}})^{T}Q_{i,j}x^{\mathbb{B}_{i,j}}$ is consistent with the joint correlative sparsity graph pattern. Recall that the sparsity graph of $Q_{i,j}$ is defined to have the edge set
\begin{equation*}
    \mathcal{E}_{i,j} := \bigcup_{k \in \mathcal{N}_{i,j}} \{ (\beta, \gamma) \in \mathbb{B}_{i,j} \times \mathbb{B}_{i,j} \: : \: \mathrm{nnz}(\beta + \gamma) \subseteq \mathcal{J}_{k} \}.
\end{equation*}
In this case the only clique where $k \in \mathcal{N}_{i,j}$ is the clique corresponding to the layer that represents that layer and the proceeding layer and hence only overlaps with a single clique such that
\begin{equation*}
    \mathcal{E}_{i,j} :=  \{ (\beta, \gamma) \in \mathbb{B}_{i,j} \times \mathbb{B}_{i,j} \: : \: \mathrm{nnz}(\beta + \gamma) \subseteq \mathcal{J}_{\mathcal{N}_{i,j}} \}.
\end{equation*}
Each $Q_{i,j}$ will then only be a function of the variables in the $i^{th}$ and $(i-1)^{th}$ layer, which matches the cliques. The monomial vectors for multipliers for relaxation order $\omega = 2$ in the cliques $\mathcal{J}_{1}, \mathcal{J}_{2} , \mathcal{J}_{3}, \mathcal{J}_{4}$ are
\begin{equation*}
    [x_{0},x_{1},1], \: [x_{1},x_{2},1], \: [x_{2},x_{3},1], \: [x_{3},x_{4},1],
\end{equation*}
respectively. This simple example has shown how the chordal structure can be preserved when introducing multipliers; we now generalise this example to any feed-forward neural network to show that the sparsity is still preserved.

\subsection{General Neural Network}
To go beyond the simple example in Section \ref{subsec:NNexample}, we introduce new notation. In particular, $g_{i,j,k}$ represents the $k^{\textrm{th}}$ inequality constraint of the $j^{\textrm{th}}$ node in the $i^{\textrm{th}}$ layer, $h_{i,j,k}$ has the equivalent meaning for equality constraints. Each node is now represented by $x_{j}^{i}$, where $i$ is the layer index and $j$ is the node index in that layer. 

For a neural network of any size there can be more than one input, hence there will be $2n_{u}$ input constraints, which can be written as:
\begin{equation*}
    g_{0,j,1}(x_{j}^{0}) = x_{j}^{0} - \underbar{$u$}_{j} \geq 0, \: g_{0,j,2}(x_{j}^{0}) = -x_{j}^{0} + \overline{u}_{j} \geq 0, \: \forall j = 1, \dots , n_{u}.
\end{equation*}

We note that often these constraints are in the form $g_{i,j,k} = g_{i,j,k}(x_{j}^{i}, x_{1}^{i-1}, \dots , x_{n_{i-1}}^{i-1})$, meaning that each constraint is only a function of the respective node and all of the variables in the previous layer. Therefore, the connection between $x_{j}^{i}$ and $x_{j^{'}}^{i-1}$ is established due to the $g_{i,j,k}$ constraint, where $j^{'}$ is any node in the $(i-1)^{\textrm{th}}$ layer. If we follow this argument for all of the nodes in the network then the joint correlative sparsity graph consists of all the nodes in each layer being joined to the neighbouring layers: this is essentially the structure of the neural network equations. 

\begin{figure}[h!] 
    \centering 
    \includegraphics[height=7cm]{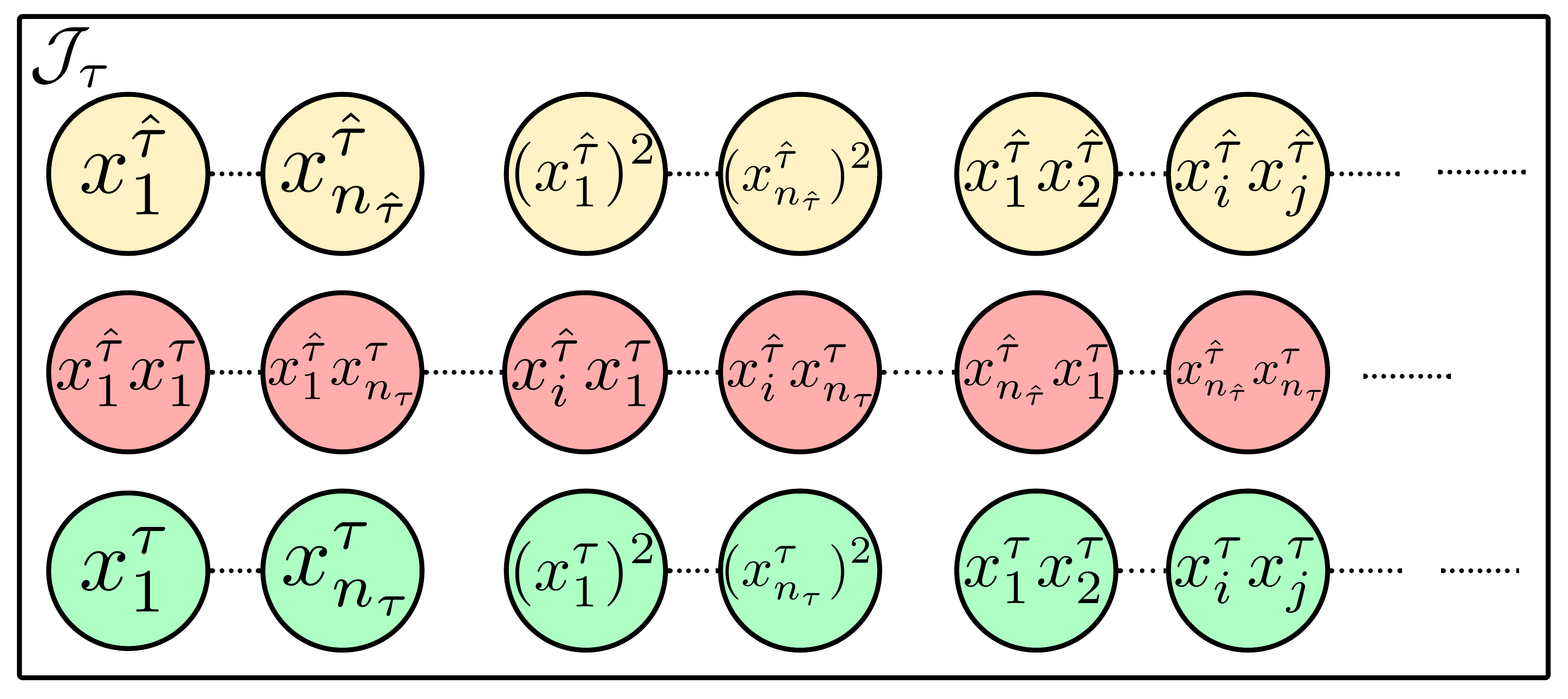}
    \caption{Diagram showing the variables that are contained in a generic clique $\mathcal{J}_{\tau}$. The yellow, green and red nodes are the monomials from layer $\hat{\tau}$, the next layer $\tau = \hat{\tau} + 1$ and overlapping terms respectively.} \label{fig:NNclique}
\end{figure}
\begin{figure}[h!] 
    \centering 
    \includegraphics[height=11cm]{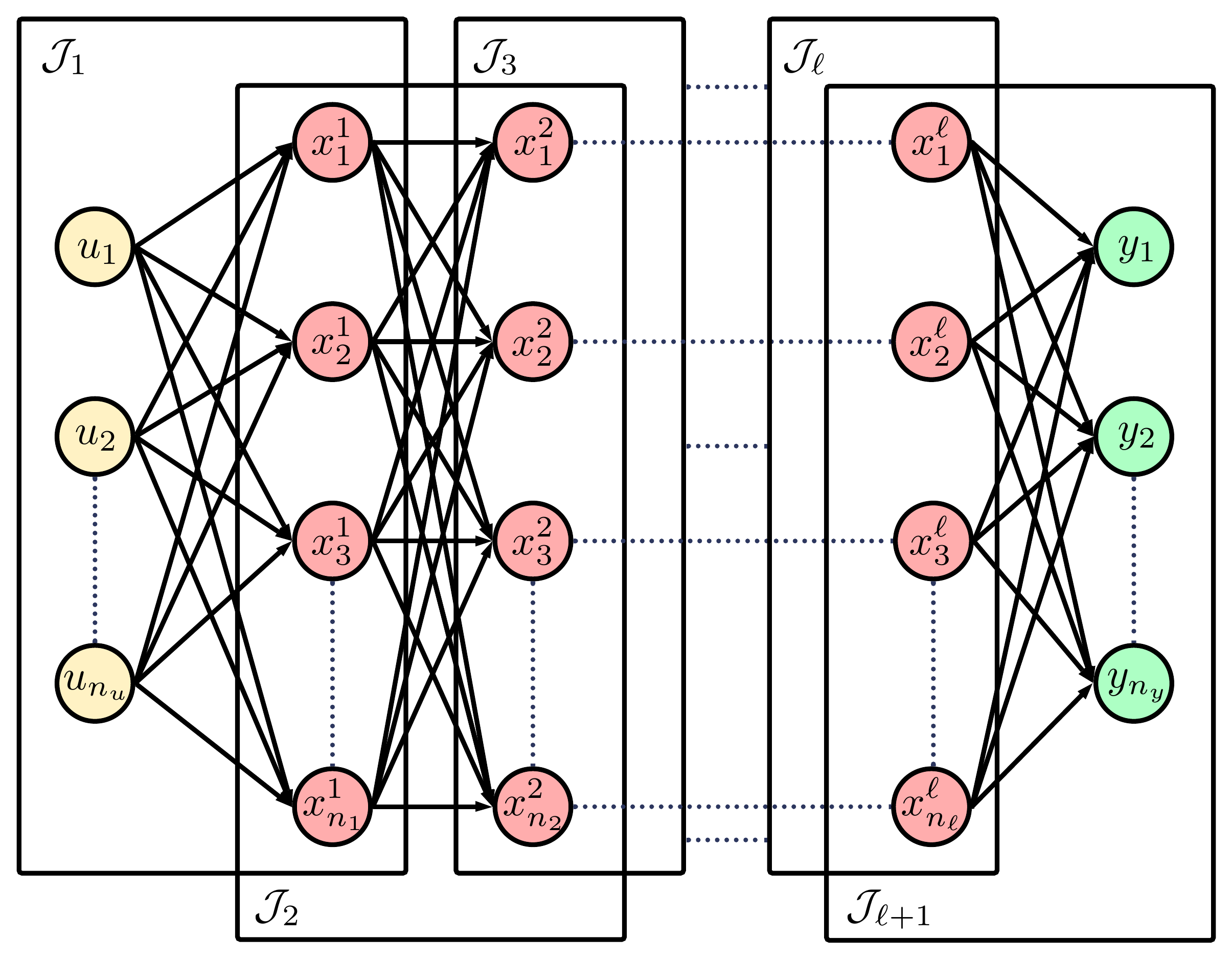}
    \caption{Diagram of a general neural network showing how the cliques are formed. The yellow nodes are the input nodes, the red nodes are the hidden layers and the green nodes are the output nodes.} \label{fig:bigcliques}
\end{figure}

The maximal cliques of the neural network verification problem consist of the overlapping layers such that for a neural network with $\ell$ layers, the verification problem can be split into $\ell + 1$ maximal cliques. This is similar as in the example case in Section \ref{subsec:NNexample} but instead there are multiple nodes in each layer, however this does not greatly impact the sparsity graph of the multiplier $Q_{i,j,k}$. The joint correlative sparsity graph for a general neural network is shown in Figures \ref{fig:NNclique} and  \ref{fig:bigcliques}. Note that as the number of layers in the neural network increases then so do the number of maximal cliques and hence the number of positive semidefinite constraints increases. As the number of nodes in each layer becomes larger then the size of these positive semidefinite constraints increases too. 

We can now use the joint correlative sparsity graph and the constraints to form the Psatz condition to verify a general neural network. For now we will only consider the Psatz where no inequality constraints are multiplied together, hence the SOS condition within the optimisation problem is:
\begin{gather*} 
    -c_{m}^{T}y + \gamma_{m} - \sum_{i=0}^{\ell} \sum_{j=1}^{n_{i}} \sum_{k=1}^{p_{i,j}} t_{i,j,k}h_{i,j,k} - \sum_{i=0}^{\ell} \sum_{j=1}^{n_{i}} \sum_{k=1}^{q_{i,j}} s_{i,j,k}g_{i,j,k} \: \mathrm{is \: (CS-)TSSOS}, \\
    s_{i,j,k} \: \mathrm{is \: SOS},\: \forall \: i = 0, \dots, \ell, \: j = 1, \dots, n_{i}, \: k = 1, \dots, p_{i,j} \\
     t_{i,j,k} \in \mathbb{R}[x], \:  \forall \: i = 0, \dots, \ell, \: j = 1, \dots, n_{i}, \: k = 1, \dots, q_{i,j}
\end{gather*}
where $p_{i,j}$ and $q_{i,j}$ are the number of inequality and equality constraints in the $i^{\textrm{th}}$ layer and $j^{\textrm{th}}$ node respectively. The multipliers $s_{i,j,k}$ and $h_{i,j,k}$ are determined by $(x^{\mathbb{B}_{i,j,k}})^{T}Q_{i,j,k}x^{\mathbb{B}_{i,j,k}}$ from the joint correlative sparsity graph.

\section{Experimental Results} \label{sec:results}
Now we examine how our approach performs in experiments. There are two main aspects to focus on, the first is how this method can improve computational time against an increasing neural network size and the second is how the accuracy of the bounds can be tightened.

All experiments were run on a 4-core Intel Xeon processor @3.50GHz with 16GB of RAM. We refer to our method as `NNSparsePsatz', which is built upon the method `NNPsatz'. We implement our method using MATLAB to create the neural network parameters, which are randomly generated from a Gaussian distribution. These parameters are then parsed into Julia where the semi-algebraic constraint set is constructed. We implement the optimisation problem with the CS-TSSOS hierarchy \cite{jwan} using the TSSOS Julia package \cite{tssos}, which constructs the SDP constraints and parses it to the SDP solver MOSEK \cite{mosek}. To show the trade-off between computational time and accuracy that is possible with this approach, we compare the results when setting the order of the multipliers to second order polynomials against setting them to their minimum level which usually sets their order to zero.  

We compare the sparse method with NNPsatz \cite{mnew2}, the equivalent method which does not exploit sparsity and is implemented with SOSTOOLS in MATLAB and also with MOSEK to solve the SDP. We also compare these results to the MATLAB package DeepSDP \cite{mfaz} with MOSEK, which is a comparable method for ReLU activation functions.

\subsection{Scalability Comparison}
There are two main ways that we can assess the scalability of the neural network verification problem. One approach is to vary the number of layers in the network and the other is to vary the number of nodes in each layer. To compare the different techniques we will consider each approach separately. For all results, we increase the number of layers or nodes until the computational time to solve the SDP exceeds 1000 seconds. 

\subsubsection{Layer Test}
\begin{figure}[h!] 
    \centering
    \includegraphics[height=8cm]{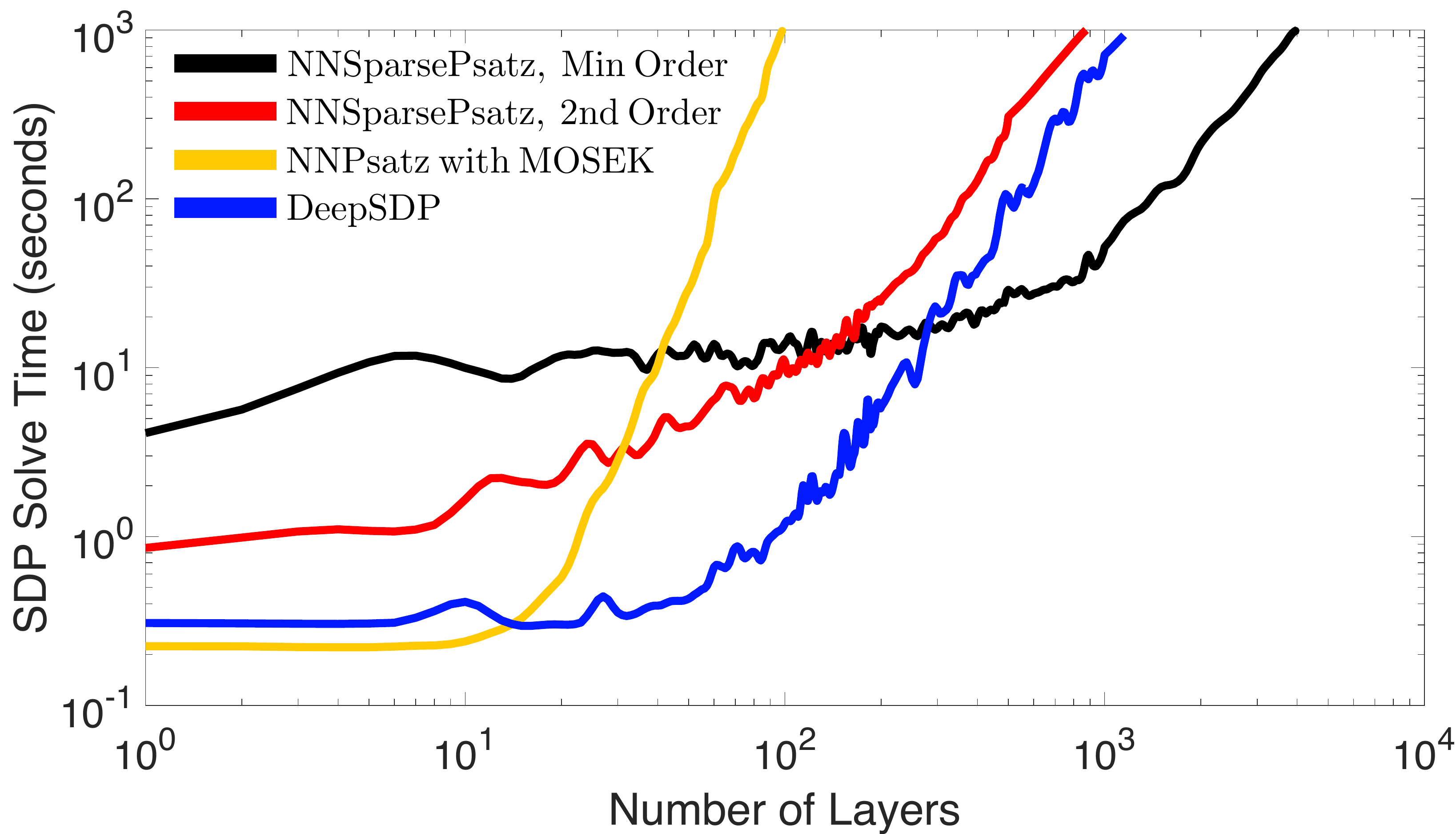}
    \caption{Comparing the computational time of different approaches for a neural network with two nodes in each layer and ReLU activation functions.} \label{fig:2layersrelu}
\end{figure}

\begin{figure}[h!] 
    \centering
    \includegraphics[height=8cm]{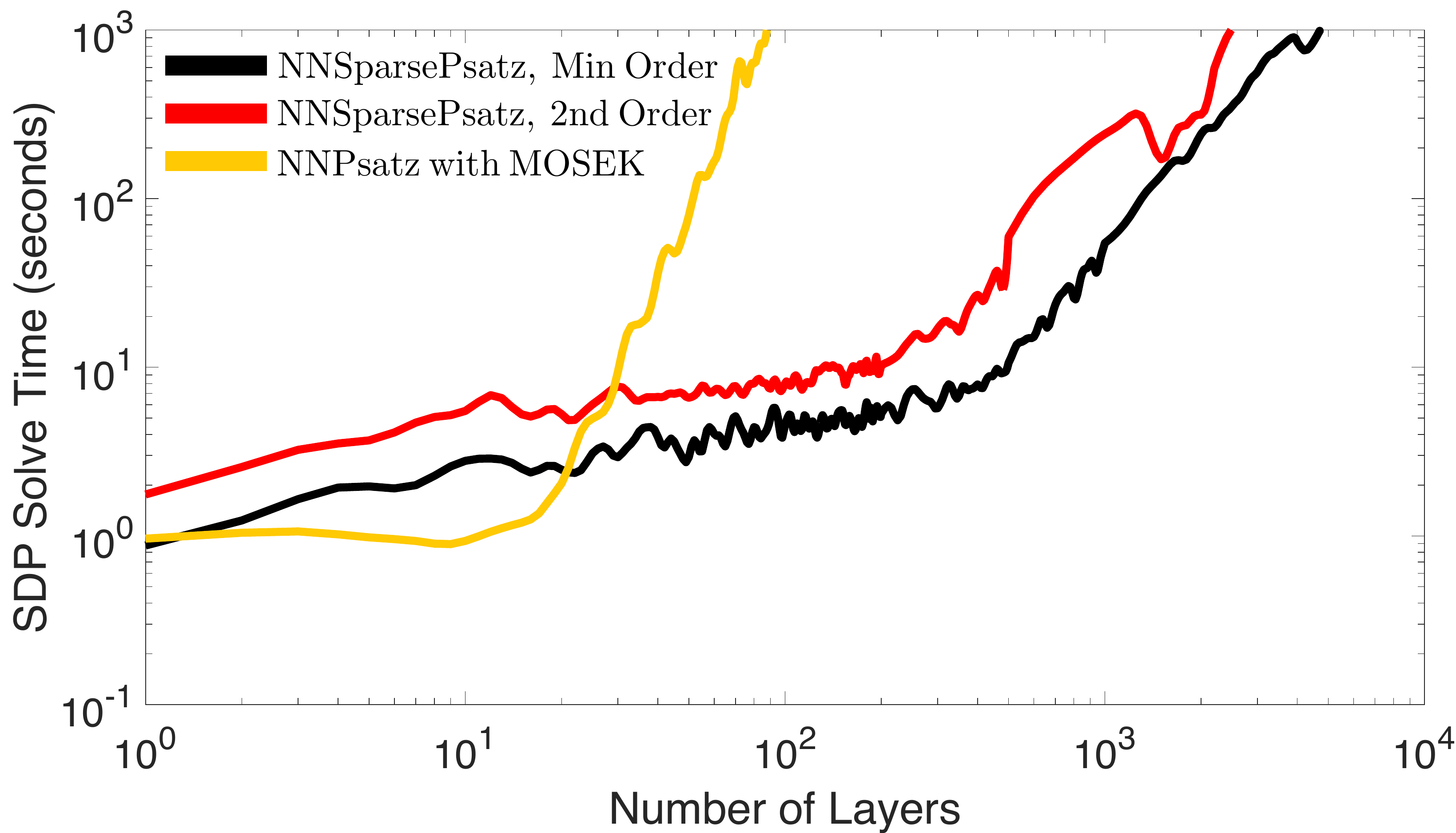}
    \caption{Comparing the computational time of different approaches for a neural network with two nodes in each layer and sigmoid activation functions.} \label{fig:2layerssig}
\end{figure}

\begin{figure}[h!] 
    \centering
    \includegraphics[height=8cm]{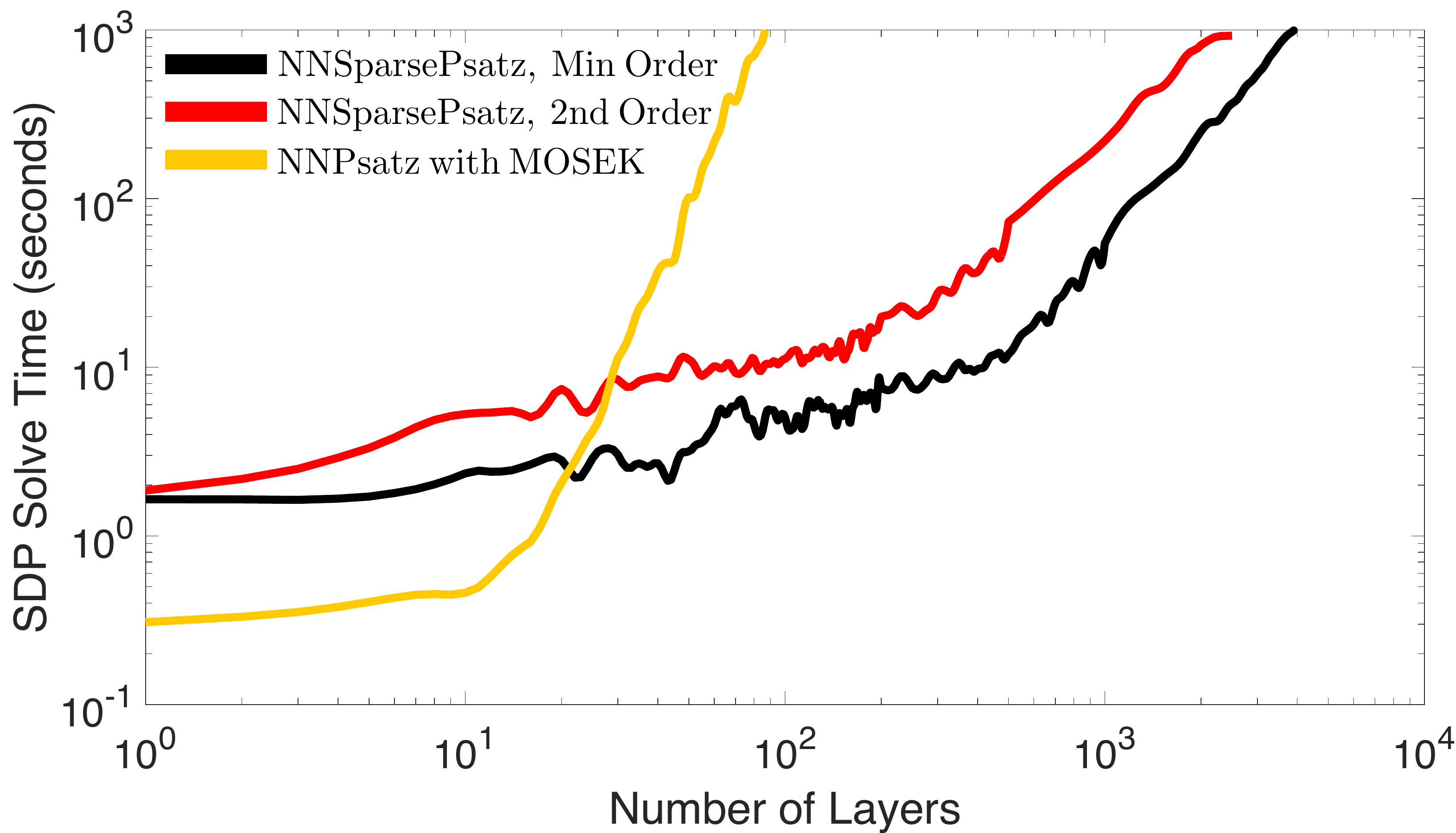}
    \caption{Comparing the computational time of different approaches for a neural network with two nodes in each layer and tanh activation functions.} \label{fig:2layerstanh}
\end{figure}

We will test a simple example to demonstrate the scalability. Consider a single input, single output neural network with ReLU activation functions, each layer will contain two nodes and we will change the size of the neural network by increasing the number of layers in the network. We show the results for ReLU, sigmoid and tanh activation functions in Figures \ref{fig:2layersrelu}, \ref{fig:2layerssig} and \ref{fig:2layerstanh} respectively. It is shown that the NNSparsePsatz with minimum order provides significantly better results when compared with DeepSDP and NNPsatz. When the order of the multipliers is set to two, the approach is more scalable than NNPsatz but less so than DeepSDP but can provide more accurate results. 

 \clearpage

\begin{figure}[h] 
    \centering
    \includegraphics[height=8cm]{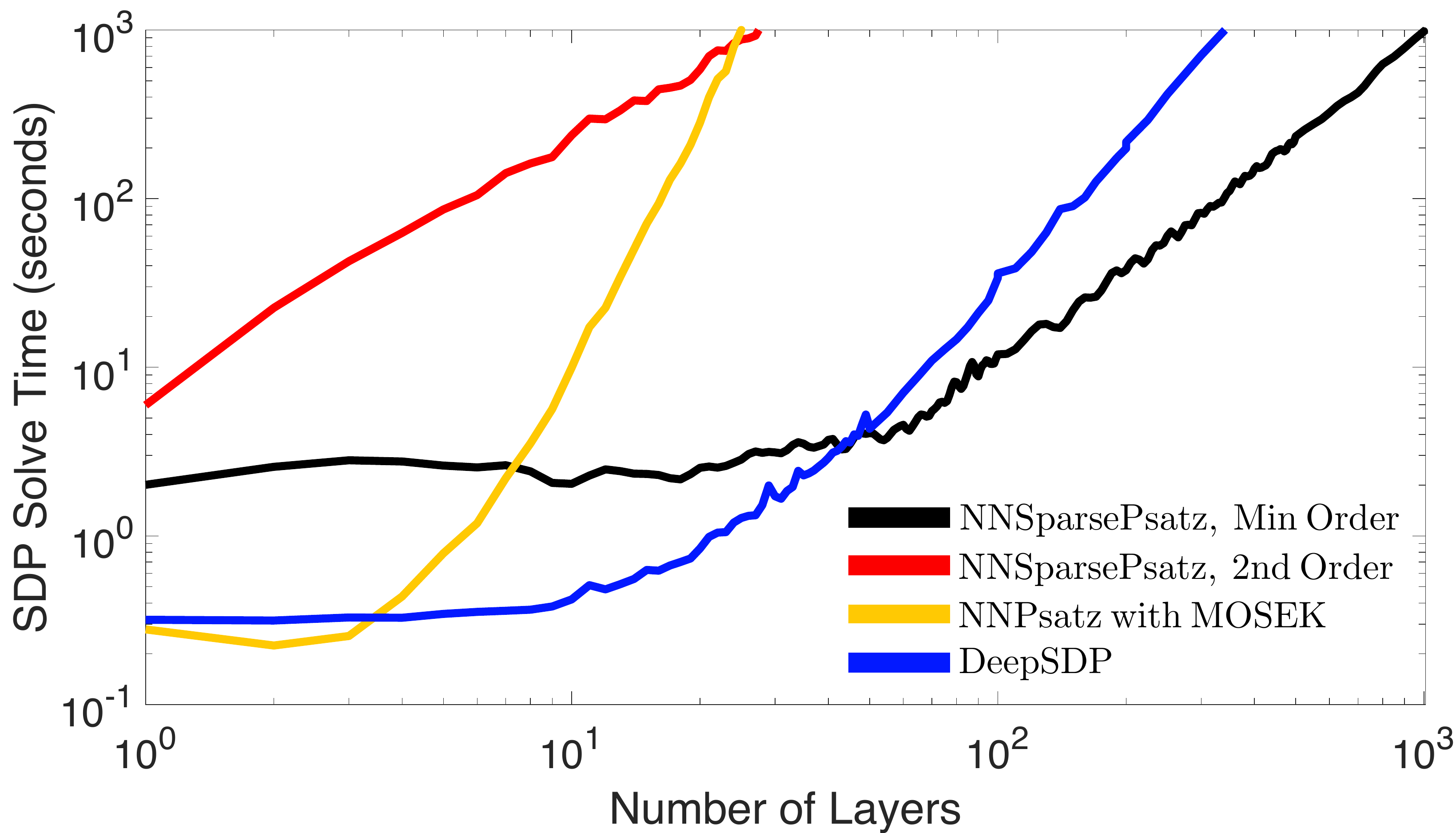}
    \caption{Comparing the computational time of different approaches for a neural network with eight nodes in each layer and ReLU activation functions.} \label{fig:8layersrelu}
\end{figure}

\begin{figure}[h] 
    \centering
    \includegraphics[height=8cm]{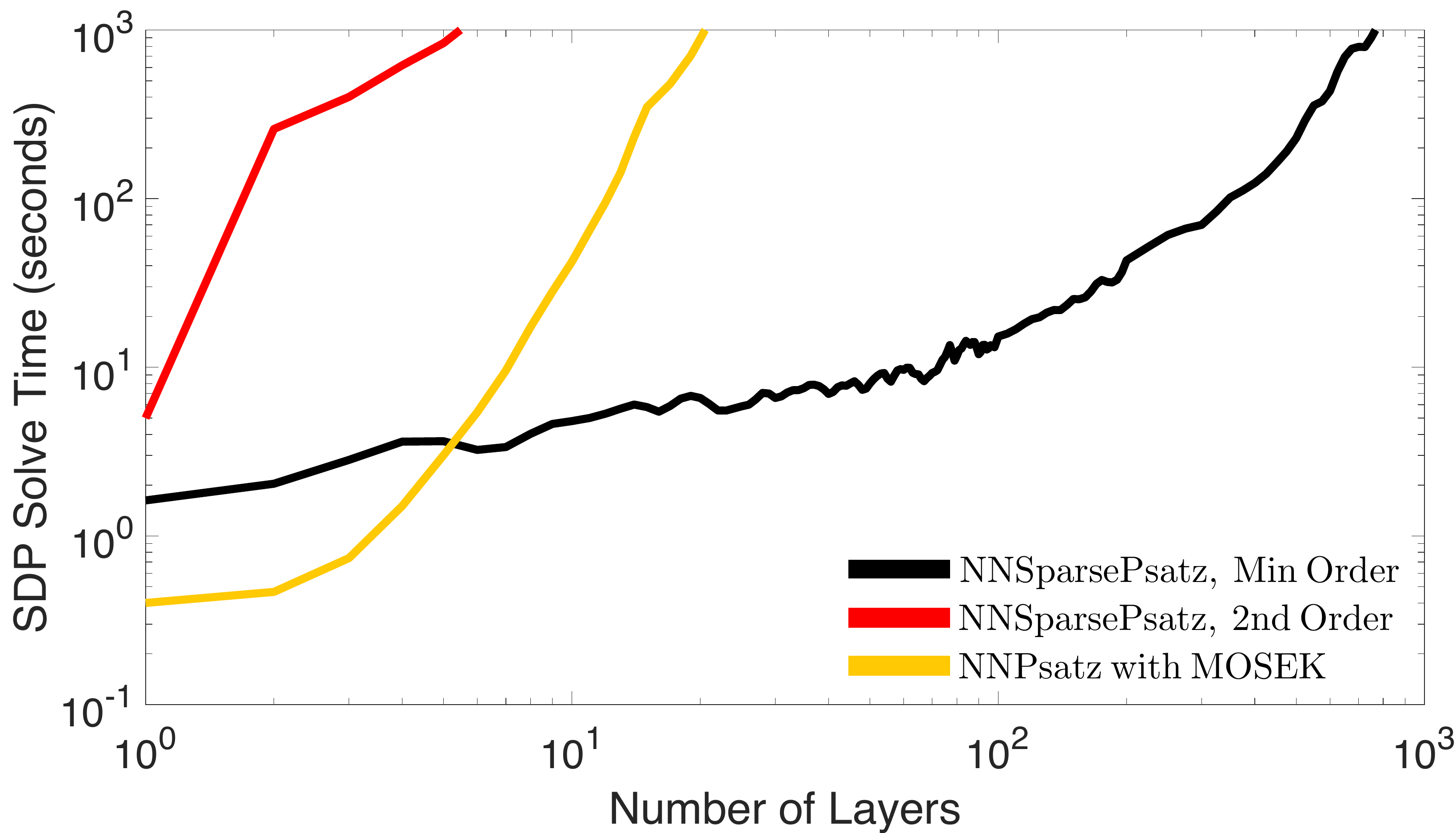}
    \caption{Comparing the computational time of different approaches for a neural network with eight nodes in each layer and sigmoid activation functions.} \label{fig:8layerssig}
\end{figure}

\begin{figure}[h] 
    \centering
    \includegraphics[height=8cm]{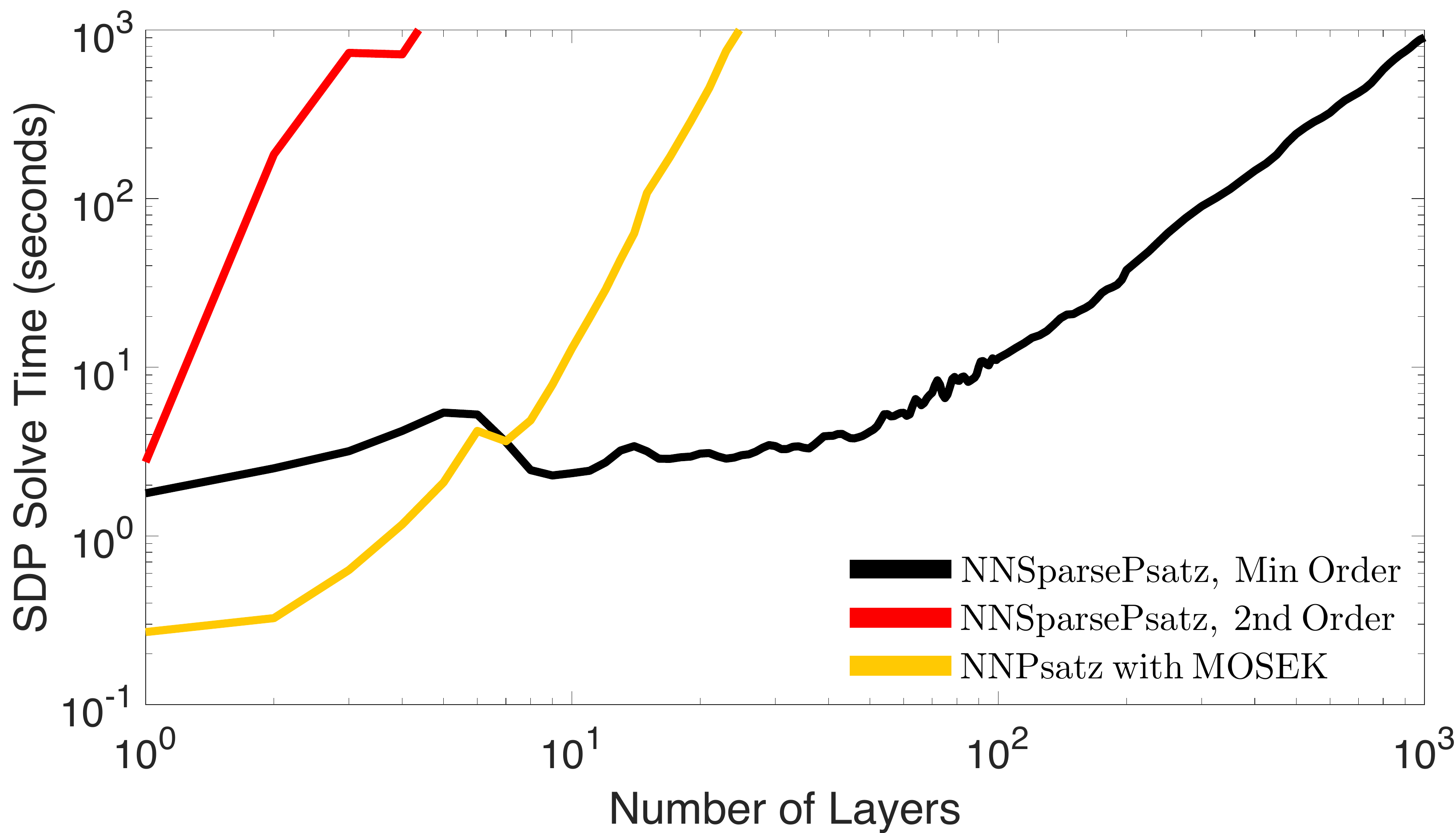}
    \caption{Comparing the computational time of different approaches for a neural network with eight nodes in each layer and tanh activation functions.} \label{fig:8layerstanh}
\end{figure}

We now increase the size of the layers to eight nodes in each layer, the remaining neural network parameters are the same as the two node example. Since there are more nodes in each layer and the size of each clique increases, the problem does not scale as well as the previous example, however there is still a significant improvement over other methods as we can see in Figures \ref{fig:8layersrelu}, \ref{fig:8layerssig} and \ref{fig:8layerstanh}. We see that for the ReLU case, NNSparsePsatz with 2nd order multipliers scales poorly, however with minimum order it scales significantly better then NNPsatz. DeepSDP in this example performs well initially, however it scales worse than NNSparsePsatz. For the sigmoid and tanh activation functions, NNSparsePsatz scales significantly better than NNPsatz, however with 2nd order multipliers it scales worse.

 \clearpage

\subsubsection{Node Test}
\begin{figure}[h] 
    \centering
    \includegraphics[height=8cm]{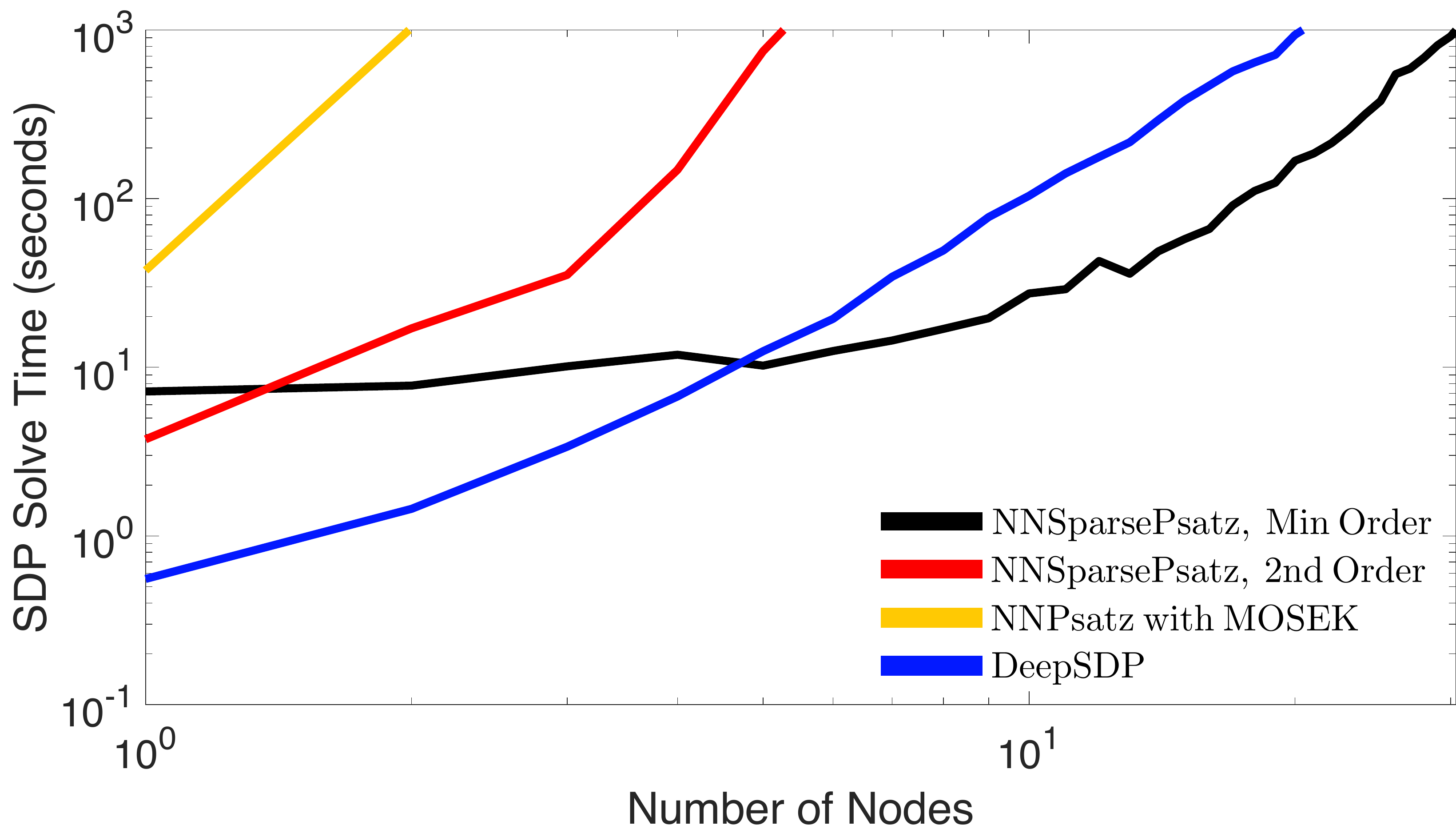}
    \caption{Comparing the computational time of different approaches for a neural network with 100 layers and ReLU activation functions.} \label{fig:5noderelu}
\end{figure}

\begin{figure}[h] 
    \centering
    \includegraphics[height=8cm]{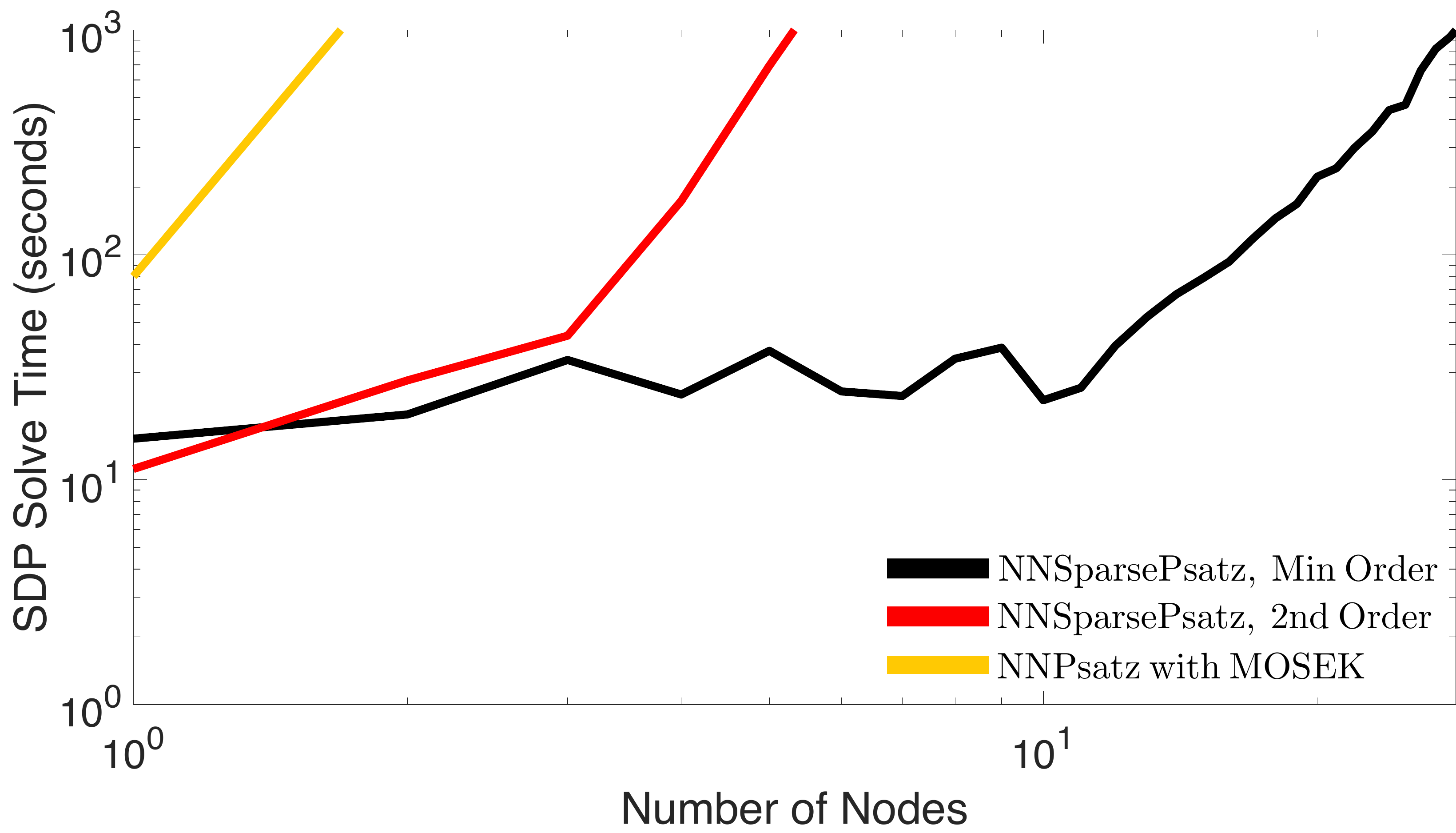}
    \caption{Comparing the computational time of different approaches for a neural network with 100 layers and sigmoid activation functions.} \label{fig:5nodesig}
\end{figure}

\begin{figure}[h] 
    \centering
    \includegraphics[height=8cm]{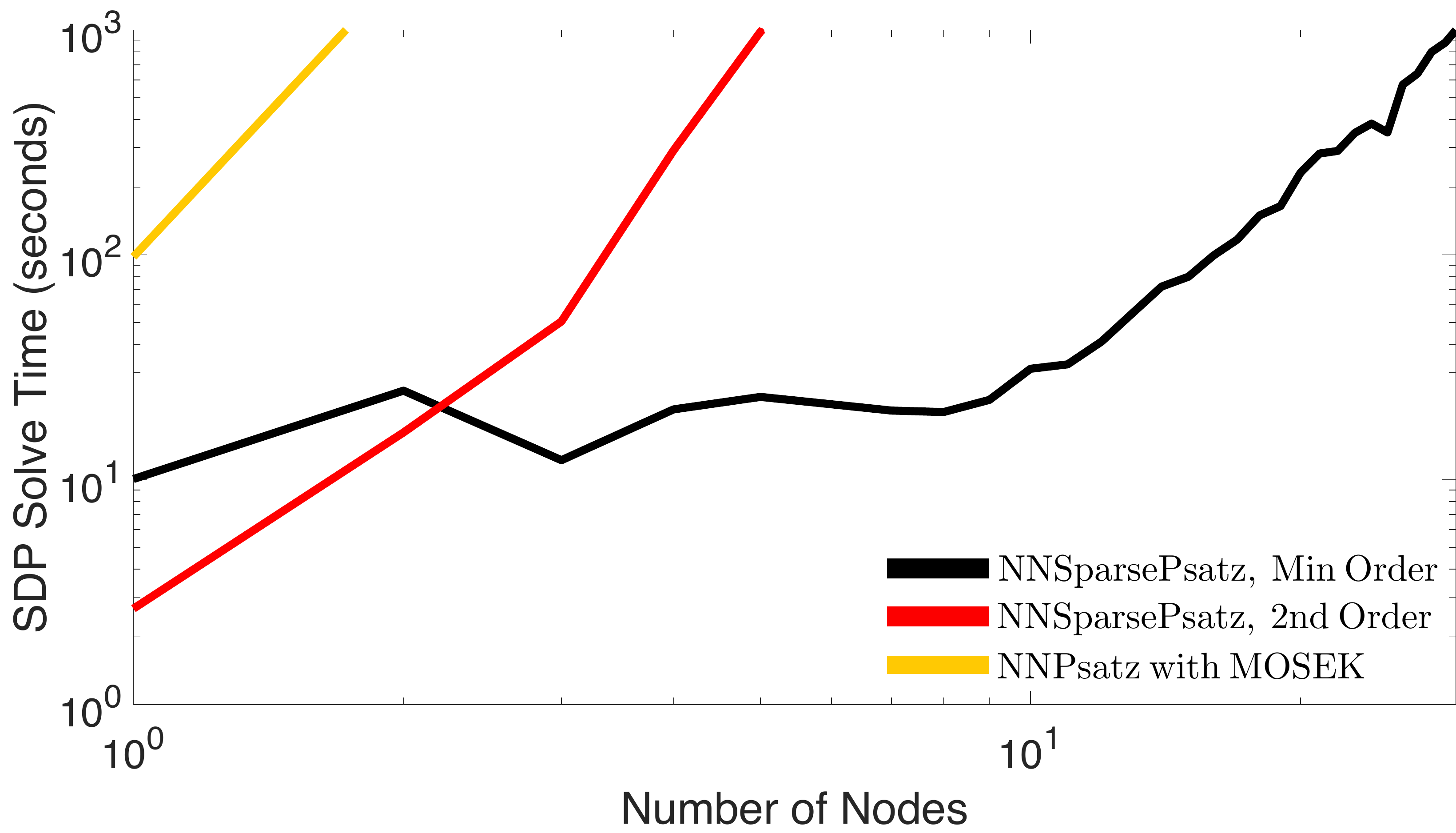}
    \caption{Comparing the computational time of different approaches for a neural network with 100 layers and tanh activation functions.} \label{fig:5nodetanh}
\end{figure}

The previous examples show the scalability is improved as we increase the number of layers in the network, we now see what happens when we vary the size of each layer. The number of layers will be fixed to 100 and the number of nodes in each of these layers will be increased. The methods are compared for ReLU, sigmoid and tanh functions and are shown in Figures \ref{fig:5noderelu}, \ref{fig:5nodesig} and \ref{fig:5nodetanh} respectively. For the ReLU activation function, NNSparsePsatz scales better than DeepSDP, whereas NNPsatz exceeds the solve time limit after only two nodes. Similarly in the sigmoid and tanh case, there is a significant improvement using NNSparsePsatz over NNPsatz.
\clearpage

\clearpage

\subsection{Accuracy Comparison}
\begin{figure}[h] 
    \centering
    \includegraphics[height=8cm]{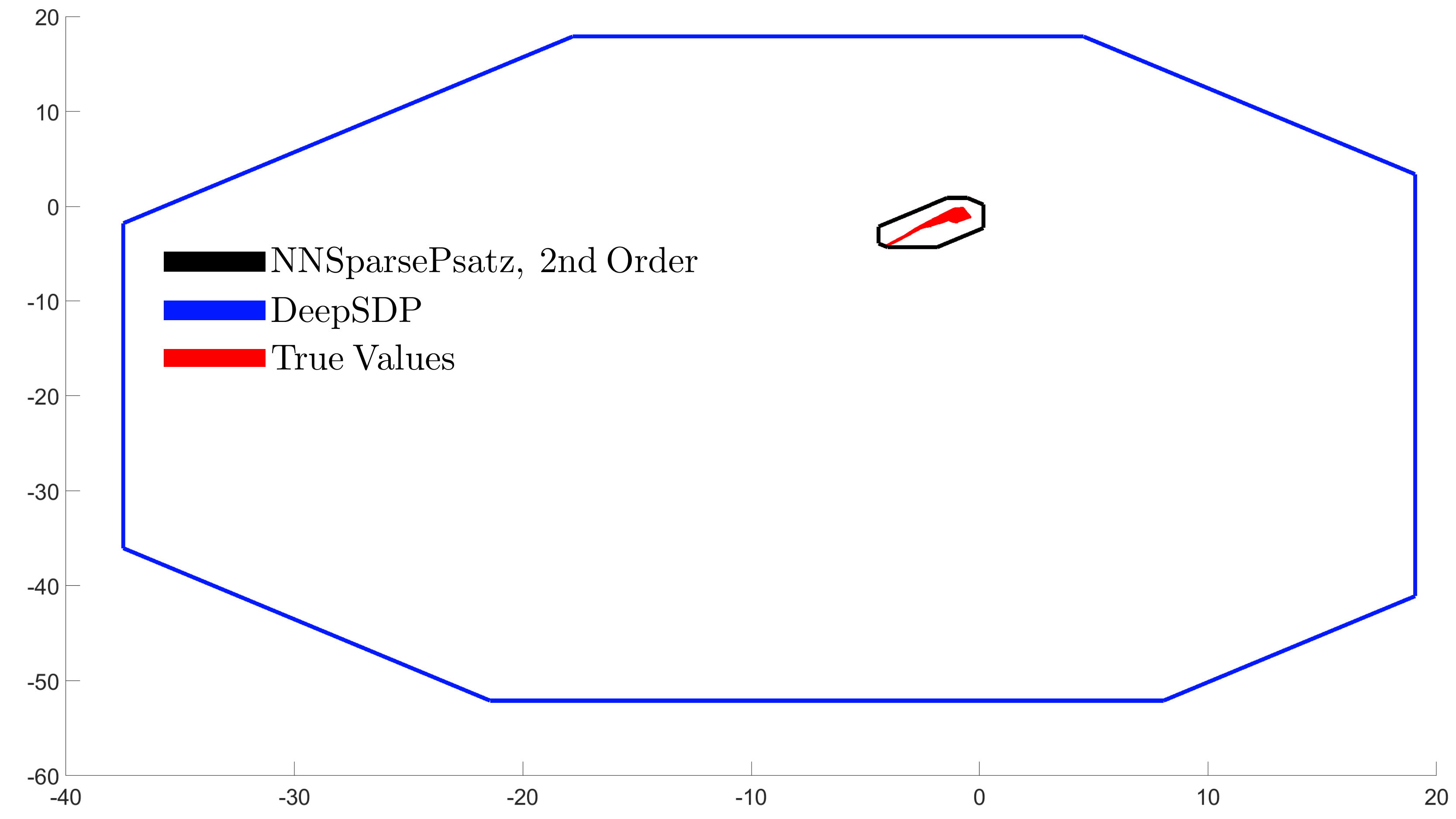}
    \caption{Comparing the accuracy of NNSparsePsatz and DeepSDP against the true values of a neural network with ReLU activation functions.} \label{fig:2Dexamplerelu}
\end{figure}

Having showed that the CS-TSSOS hierarchy can improve the scalability of the problem, we now turn our attention to the accuracy of the methods. We consider a two input/output neural network with eight layers and eight nodes in each layer and ReLU activation functions. The input space is set to [5,15]. It is shown in Figure \ref{fig:2Dexamplerelu} that the NNSparsePsatz method with second order polynomials greatly improves the tightness of the bounds on the output set to the point where they are near optimal. This approach is more computationally expensive over DeepSDP, however the accuracy improvement is significant. If we instead used minimum order multipliers in NNSparsePsatz, then the accuracy would be the same as DeepSDP.

\begin{figure}[h] 
    \centering
    \includegraphics[height=8cm]{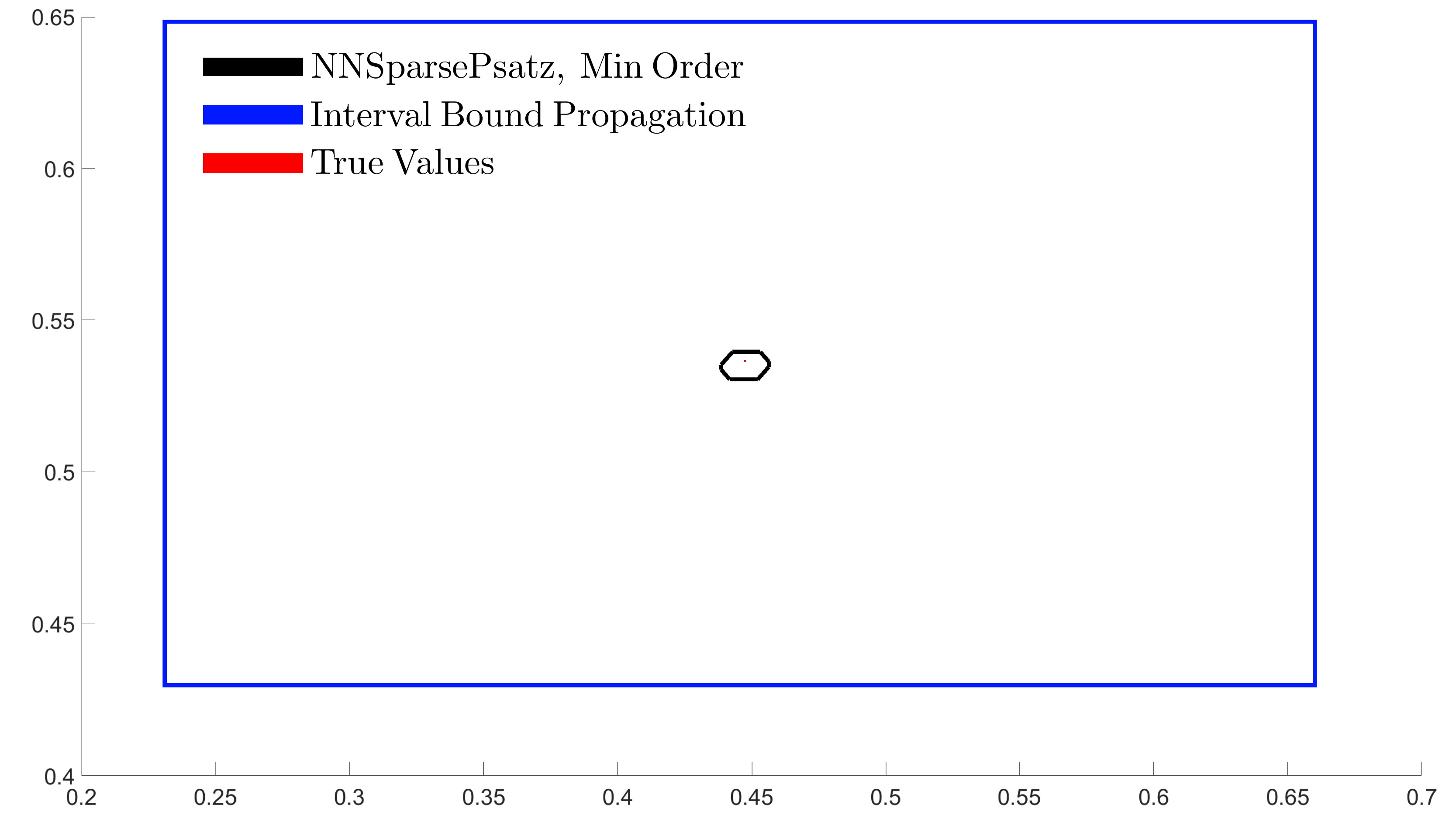}
    \caption{Comparing the accuracy of NNSparsePsatz and Interval Bound Propagation against the true values of a neural network with sigmoid activation functions.} \label{fig:2Dexamplesigmoid}
\end{figure}

\begin{figure}[h] 
    \centering
    \includegraphics[height=8cm]{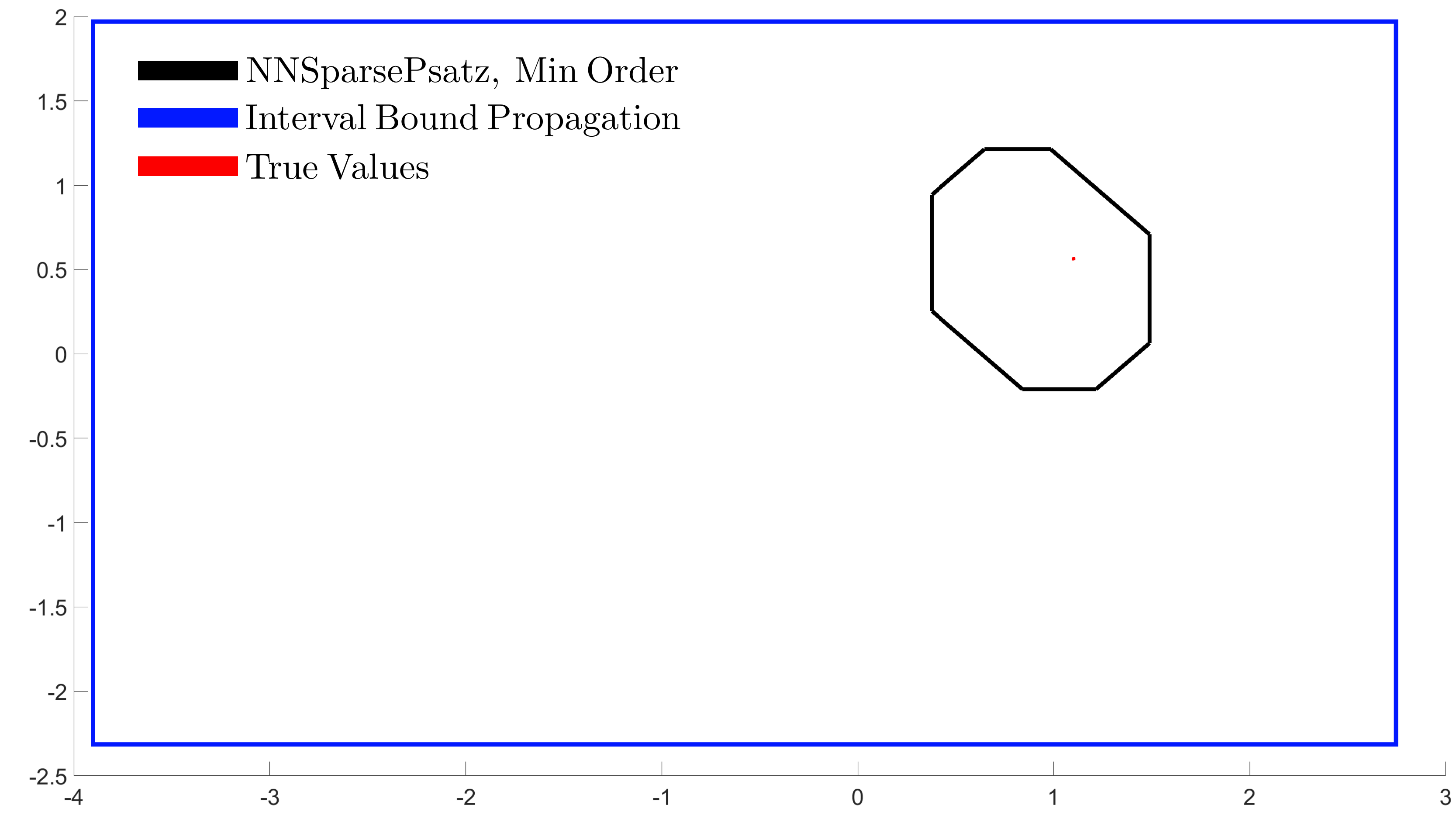}
    \caption{Comparing the accuracy of NNSparsePsatz and Interval Bound Propagation against the true values of a neural network with tanh activation functions.} \label{fig:2Dexampletanh}
\end{figure}

For the sigmoid activation function we consider a ten layer network with fifty nodes in each layer, again with two inputs and two outputs. DeepSDP cannot be used, so instead we use interval bound propagation, which is very conservative. As shown in Figure \ref{fig:2Dexamplesigmoid} the output of the neural network converges to a point and NNSparsePsatz is able to obtain tight bounds on the output. If we tried to optimise this problem using NNPsatz it would be intractable. We see similar results in the case of the tanh activation function (Figure \ref{fig:2Dexampletanh}); for this network we choose a twelve layer network with fives nodes in each layer.


\section{Conclusion} \label{sec:conc}
In this paper, we propose a framework to address the neural network verification problem, through placing bounds on the non-linear activation functions. Using a theory called the Postivstellensatz we are able to trade-off solution accuracy with computational time within the optimisation problem. We show that the semi-algebraic set that is constructed possesses a significant amount of sparsity and hence the scalability of this method can be improved by using ideas from sparse polynomial optimisation. Most constraints that are used in problems of this type will exhibit a sparsity pattern, which essentially comes from the natural cascading structure of the neural network. We then implement the optimisation framework using the CS-TSSOS hierarchy and show that it both improves the computation time to solve the resulting SDP against similar methods and can do so by tightening the bounds on the neural network outputs. 

This paper leaves scope for future work. First, we can use these ideas in feedback control systems and combining them with notions of stability. Since we used random neural networks in this paper to show scalability, it would be interesting to see how effective these methods are in real neural network examples. Another way of improving the scalability of the neural network verification problem is to use neural network pruning \cite{dbla} to reduce the size of the neural network and hence make it more efficient to verify. Related works in \cite{rdrum} synthesise a reduced order neural network with robustness guarantees. It may also be possible to combine the approaches used in this paper with boosting methods. Finally, the activation functions used in this paper are ReLU, sigmoid and tanh; it would be interesting to see the performance of the ideas from this paper when applied to other activation functions. There are also other properties of neural networks that can quantify robustness, which would benefit from the approach of this paper. With a combination of these improvements, neural networks can be verified more effectively in the future. 

\printbibliography

\end{document}